\documentclass[preprint,review,12pt]{elsarticle}

\journal{Information and Software Technology}
\usepackage{amsmath,amssymb,amsfonts}
\usepackage{algorithmic}
\usepackage{graphicx}
\usepackage{subcaption}
\usepackage{textcomp}
\usepackage{xcolor}
\usepackage[most]{tcolorbox}
\usepackage{fontawesome}
\usepackage{multirow}
\usepackage{enumitem}
\usepackage{booktabs}
\usepackage{url}
\usepackage{soul}
\usepackage{xcolor}
\usepackage{colortbl}

\usepackage{pdflscape,longtable,booktabs,array,ragged2e,calc,makecell}
\definecolor{darkgreen}{rgb}{0.0, 0.0, 0.0}
\definecolor{darkblue}{rgb}{0.0, 0.0, 0.0}
\definecolor{r2added}{HTML}{000000}
\definecolor{r2revised}{HTML}{000000}
\newcommand{\radd}[1]{#1}
\newcommand{\rrev}[1]{#1}
\definecolor{r3added}{HTML}{000000}
\definecolor{r3revised}{HTML}{000000}
\newcommand{\rthreeadd}[1]{#1}
\newcommand{\rthreerev}[1]{#1}

\definecolor{greyb}{RGB}{248, 248, 248}
\definecolor{greycon}{RGB}{242, 242, 242}
\definecolor{greya}{RGB}{240, 240, 240}
\definecolor{greyquote}{RGB}{245, 245, 245}
\definecolor{dgrey}{RGB}{224, 224, 224}
\def\mybarhhigh#1#2{%
   {\color{black}\rule{#1mm}{4pt}}  #2}

\definecolor{signif}{RGB}{228, 245, 246}
\definecolor{signifl}{RGB}{229, 229, 200}
\definecolor{experience}{RGB}{236, 233, 237}
\definecolor{health}{RGB}{222, 232, 252}

\newcommand{\vcenteredincludep}[1]{\begingroup
\setbox0=\hbox{\includegraphics[height=1.1em]{#1}}%
\parbox{\wd0}{\box0}\endgroup}

\newcommand{\todo}[1]{}
\renewcommand{\todo}[1]{{\color{red} TODO: {#1}}}
\newcommand{\hquote}[1]{\,\faCommentO\,\textit{#1}\,}

\newcommand{\hconcep}[1]{\sethlcolor{greycon}\hl{#1}}
\newcommand{\hcate}[1]{\sethlcolor{dgrey}\hl{#1}}

\newcommand{\memo}[1]{
\vspace{0.5em}
\hspace{-1.5em}
        \tcbox[on line,
        boxsep=3pt, left=10pt,right=10pt,top=2pt,bottom=2
        pt,
        colframe=white,colback=gray!10]{%
            \parbox{0.9\linewidth}{%
            \vspace{0.5em}
            \faStickyNoteO \hspace{0.3em}#1
                \vspace{0.5em}
            }
        }\vspace{0.5em}
}%

\def\BibTeX{{\rm B\kern-.05em{\sc i\kern-.025em b}\kern-.08em
    T\kern-.1667em\lower.7ex\hbox{E}\kern-.125emX}}
\makeatletter
\def\ps@pprintTitle{\let\@oddhead\@empty\let\@evenhead\@empty\def\@oddfoot{\hfil\thepage\hfil}\let\@evenfoot\@oddfoot}
\makeatother
\begin{document}
\begin{frontmatter}

\title{ChroniUXMag: A Persona-Driven Framework for Inclusive mHealth Requirements Engineering}

\author[Monash]{Wei Wang}
\cortext[cor1]{Corresponding author}
\ead{wei.wang7@monash.edu}

\author[Pal] {Devi Karolita\corref{cor1}}
\ead{devikarolita@it.upr.ac.id}

\author[Deakin]{Hourieh Khalajzadeh}
\ead{hkhalajzadeh@deakin.edu.au }

\author[Monash]{John Grundy}
\ead{john.grundy@monash.edu}

\author[Deakin]{Anuradha Madugalla}
\ead{anuradha.madugalla@deakin.edu.au}

\author[Telstra]{Humphrey O. Obie}
\ead{humphreyobie@gmail.com}

\affiliation[Monash]{organization={Monash University},%
            city={Melbourne},
            country={Australia}}

\affiliation[Pal]{organization={Palangka Raya University},%
            city={Palangka Raya},
            country={Indonesia}}

\affiliation[Deakin]{organization={Deakin University},%
            city={Melbourne},
            country={Australia}}

\affiliation[Telstra]{organization={Telstra Health},%
            city={Melbourne},
            country={Australia}}

\begin{abstract}
\textbf{Context:} Mobile health (mHealth) applications are increasingly adopted for chronic disease management, yet \rrev{people living with chronic diseases continue to face challenges related to accessibility, inclusivity, usability, and sustained engagement with these applications}. Although contemporary requirements engineering (RE) practices can accommodate changing requirements, chronic disease mHealth introduces domain-specific challenges related to clinical context, caregiver involvement, trust, and long-term use.

\noindent
\textbf{Objective:} This study introduces \textit{ChroniUXMag}, a framework for eliciting and analysing inclusivity requirements in mHealth design. Building on InclusiveMag and GenderMag principles, the framework aims to help researchers and practitioners systematically capture and evaluate factors that influence how individuals with chronic conditions perceive, trust, and interact with mHealth systems.

\noindent
\textbf{Method:} ChroniUXMag was developed using the \textit{Scope} and \textit{Derive} stages of InclusiveMag as a meta-methodological guide, while the underlying evidence was generated through a systematic literature review, focus groups, interviews, and an \rrev{online survey}. In the first stage, candidate inclusivity facets were identified from literature and empirical data. In the second stage, these facets were organised into personas representing diverse health situations, attitudes, and digital practices, and integrated into an adapted cognitive walkthrough form.

\noindent
\textbf{Results:} \rrev{The qualitative analysis generated thirteen categories, which were consolidated into six ChroniUXMag facets: health situation, health attitude, cultural differences, level of dependency, trust in mHealth apps, and users' involvement with the app.} These facets support structured, persona-driven evaluations that reveal inclusivity barriers often missed by traditional usability assessments.

\noindent
\textbf{Conclusion:} \rrev{ChroniUXMag contributes to RE research by providing an empirically grounded, facet-driven framework and persona-based artefacts for representing inclusivity concerns during requirements elicitation and early analysis. Its utility in professional RE practice remains to be evaluated with practitioners.} Future work will extend the third stage (\textit{Apply}) through practitioner-led evaluation in real-world design contexts.

\end{abstract}

\end{frontmatter}
\begin{center}\footnotesize
Author accepted manuscript. Published article: \\
\url{https://doi.org/10.1016/j.infsof.2026.108375}\\
\copyright\ 2026. This manuscript version is made available under the\\
CC-BY-NC-ND 4.0 license: \url{https://creativecommons.org/licenses/by-nc-nd/4.0/}
\end{center}

\color{darkblue}
\section{Introduction} \label{introduction}
\rrev{Chronic diseases, including diabetes, hypertension, and cardiovascular disease, are leading causes of death and disability worldwide \cite{WHO_2024}. The need for continuous monitoring has driven the development of \rthreerev{\textbf{mobile health (mHealth)} applications (apps) that support chronic-disease self-management, including medication adherence, symptom and health-status monitoring, and lifestyle management} \cite{bricca2022quality,hamine2015impact}. From a requirements engineering (RE) perspective, the central challenge is to elicit, analyse, and validate requirements for people whose capabilities, health conditions, care arrangements, and support needs differ and change over time. Although mHealth apps can support monitoring and self-management, they continue to present accessibility, inclusivity, usability, and sustained-engagement challenges \cite{kim2019mobile,van2022digital}. \rthreerev{Accessibility concerns whether people with varying capabilities can perceive and use a system, whereas usability concerns whether they can complete tasks effectively and efficiently. Inclusivity concerns whether heterogeneous users and contexts are represented in requirements work, while sustained engagement concerns whether people can continue using the system over time.}}

\color{darkblue}
\rrev{Although mHealth adoption has been studied in relation to health literacy, age, and education \cite{meyerowitz2020rates}, the requirements implications of chronic-disease variability remain insufficiently articulated. Requirements must account for changes within a person---including fluctuating symptoms, treatment burden, and caregiver involvement---as well as differences between people in trust, digital capability, and support context \cite{harvey2012future}. Addressing this variability requires a structured way to elicit and analyse inclusivity-relevant requirements.}

\color{darkblue}
\rrev{Existing digital-health work often remains clinical and treatment-centric \cite{hamine2015impact,pollack2016closing,siek2006pride}. Consequently, socio-technical factors such as trust, privacy, cultural attitudes, and changing cognitive, physical, and social contexts are not always made explicit during requirements elicitation and analysis \cite{levy2023sustaining}.}

\color{darkblue}
To address these challenges, our study adopts a facet-driven, persona-based approach for making inclusivity-relevant requirements more explicit in the mHealth domain. Inspired by prior inclusivity methods such as InclusiveMag \cite{mendez2019gendermag} and GenderMag \cite{burnett2016gendermag}, as well as their extensions including AgeMag \cite{mcintosh2021evaluating}, SESMag \cite{agarwal2023socioeconomicmag}, and Elderly HealthMag \cite{xiao2026elderly}, we developed \hconcep{Chronic Disease User Experience Magnifier} (ChroniUXMag). Existing Mag-based approaches have successfully highlighted inclusivity concerns related to demographic characteristics such as gender \cite{burnett2016gendermag}, age \cite{mcintosh2021evaluating}, and socioeconomic background \cite{burnett2021toward}, and in some cases health conditions among older adults \cite{xiao2026elderly}. However, their facets mainly emphasise relatively stable user attributes rather than changes in health condition, dependency, and care context over time. In contrast, chronic disease management introduces evolving, condition-specific, and relational dimensions, including fluctuating health status, caregiver involvement, long-term treatment adherence, and shifting dependency levels. These gaps motivate the need for a framework that captures mHealth-specific inclusivity factors in a form usable for requirements work. A persona-driven RE approach is appropriate in this context because chronic disease mHealth must represent diverse stakeholder situations, make interdependencies between patients and caregivers explicit, and support reasoning about requirement trade-offs across different health trajectories. Rather than treating InclusiveMag as a standalone framework-synthesis method, ChroniUXMag uses its meta-methodological logic to structure the generation of a domain-specific inclusivity method. Specifically, the Scope stage guided the identification of candidate inclusivity factors from literature and empirical studies, and the Derive stage guided their organisation into personas and an adapted analytic process. The goal of this study is to identify and operationalise inclusivity-relevant facets that can better support requirements elicitation and analysis for chronic disease mHealth applications. We therefore examine what factors shape inclusive requirements in chronic disease mHealth and how those factors can inform a practical framework for requirements elicitation and analysis. This study is guided by the following research question: which inclusivity-relevant factors should be represented to better support requirements elicitation and analysis in chronic disease mHealth?

\color{darkblue}
By systematically linking empirical evidence to persona development and evaluation protocols, this approach supports requirements specification with a primary focus on \textbf{inclusivity} in chronic disease management. The key contributions of this research are:
\begin{itemize}
    \item Introducing ChroniUXMag, a framework for eliciting and analysing inclusivity-relevant requirements in chronic disease mHealth.
    \item Defining a set of empirically grounded facets and personas that represent diverse health situations, support arrangements, and user contexts.
    \item Adapting a persona-driven cognitive walkthrough that embeds ChroniUXMag facets to identify both general usability issues and inclusivity concerns in mHealth apps.
\end{itemize}

\color{black}
The paper is organised as follows. Section \ref{background} summarises the related work. Section \ref{method} details the methodology we employed. Section \ref{results} presents the results of our user study. Sections 5--7 describe the derivation, operationalisation, and illustrative use of the ChroniUXMag requirements framework, respectively. Section \ref{threat to validity} describes the threats to the validity of our results and Section \ref{future work} discusses future work. Section \ref{conclusion} concludes the paper.

\color{black}
\section{Background} \label{background}

\color{black}
\subsection{Chronic disease and mHealth apps} \label{Chronic Disease}

\color{darkblue}
\rrev{The development of mHealth technologies that improve access to health information and resources is important for addressing physical and social inequities \cite{han2010professional}. One persistent challenge is non-adherence---inconsistent adherence to prescribed medication, self-monitoring, or lifestyle plans---which is common in chronic-disease management and is particularly consequential where access to ongoing care is constrained \cite{alwan2009review,sabate2003adherence}. From an RE perspective, non-adherence can be influenced by whether an app fits users' capabilities, contexts, and everyday practices. Inaccessible interfaces or inadequate contextual support can increase cognitive burden and undermine sustained use. Flexible and personalised support is therefore needed as users' circumstances evolve \cite{choe2017semi}. In this paper, \radd{accessibility denotes whether an mHealth system can be perceived, understood, and operated by people with varying physical, cognitive, and sensory capabilities; inclusivity denotes whether requirements artefacts and practices represent heterogeneous health situations, social contexts, and support arrangements; usability denotes the effectiveness and efficiency of task completion; and sustained engagement denotes users' willingness and ability to continue using the system over time.} Existing solutions nevertheless remain clinically focused or assume a single user \cite{pollack2016closing,siek2006pride,lee2024identify}. Prior accessibility work has addressed disabilities \cite{radcliffe2021pilot,zhou2020making}, visual impairments \cite{alshayban2020accessibility,milne2014accessibility}, and economic disadvantage \cite{stowell2018designing}, but these efforts do not systematically represent the broader socio-technical variability of chronic disease.}

\color{darkblue}
From a requirements perspective, this suggests that many mHealth systems are developed based on simplified or generalised user models, leading to under-specified requirements regarding fluctuating health status, multimorbidity, caregiver involvement, trust expectations, and digital literacy. Mack et al. \cite{mack2021we} highlight the limited attention given to accessibility challenges specific to chronic illness, indicating that the needs of chronically ill users are rarely formalised as structured requirements concerns. Consequently, accessibility in chronic disease mHealth remains insufficiently integrated into requirements elicitation, modelling, and validation processes.  However, this approach poses challenges due to the \textbf{diverse} and \textbf{dynamic} nature of users with chronic diseases. Diversity refers to the wide variation between individuals, including differences in disease type, symptom severity, comorbidities, age, digital literacy, cultural background, and reliance on caregivers \cite{harvey2012future}. Dynamism, in contrast, refers to changes within the same individual over time \cite{gregor2002designing,di2019chronic}. As chronic conditions progress, stabilize, or fluctuate, patients’ functional abilities, cognitive load, treatment regimens, and support needs evolve. Chronic diseases often coexist with other medical or psychological disorders, leading to increased healthcare needs and greater diversity among users \cite{lorig2003self, di2019chronic}. As a result, user requirements are neither uniform across users nor stable over time, creating unique challenges for requirements elicitation, specification, and validation in mHealth systems. Importantly, this work does not advocate a “one-app-for-all” solution; instead, it focuses on inclusive requirements engineering practices that explicitly acknowledge and manage user heterogeneity within realistic and bounded mHealth application scopes.

\color{darkblue}
Chronic disease prevalence increases with age, and older adults therefore constitute a substantial proportion of mHealth users. Ageing introduces additional requirements considerations, including age-related changes in vision, cognition, motor ability, and digital confidence \cite{charness2009aging, vaportzis2017older}. These findings align with the Technology Acceptance Model (TAM) \cite{davis1989perceived}, which provides a theoretical foundation for understanding technology adoption by emphasising the roles of usability, usefulness, and user attitudes. For older adults, additional considerations—such as age-related cognitive decline, sensory limitations, and lower digital literacy—must be factored into the design and requirements elicitation processes \cite{isakovic2016usability,bertolazzi2024barriers}.

\color{darkblue}
\subsection{RE in digital health}
Requirements Engineering (RE) research in digital health and other socio-technical domains has increasingly emphasized the need to account for human factors, stakeholder diversity, contextual use conditions, and domain-specific constraints during system design \cite{fricker2015digitalhealth}. Prior work highlights the importance of human-centred RE approaches that explicitly consider stakeholder goals, emotions, values, abilities, and everyday practices when eliciting and specifying requirements \cite{horkoff2016interactive,taveter2019method,pinto2019inclusion}. This line of work has also stressed that requirements quality depends on how well elicitation methods capture the perspectives of multiple stakeholders, including end users, clinicians, caregivers, developers, and those affected by exclusion or accessibility barriers. In the context of chronic disease management, these RE concerns are intensified by the dynamic and diverse nature of users, care arrangements, and use contexts, as well as by the partial, uncertain, and evolving character of healthcare requirements \cite{sutcliffe2020known}. Requirements must accommodate changes in health conditions, treatment routines, and support needs over time, while also addressing substantial variation across users in health status, digital literacy, dependency on caregivers, and engagement with technology. These factors introduce requirements concerns that extend beyond purely technical functionality, requiring systematic consideration of evolving and heterogeneous user needs during requirements elicitation and specification for mHealth systems. This includes accessibility-related concerns, inclusive participation of different stakeholder groups, and the need to reason about requirements that may vary across time, context, and capability. To effectively capture heterogeneous user needs, RE methods need stronger ways to represent variation across users, contexts, and support arrangements. One promising direction is the use of persona-driven approaches within broader human-centred RE practice. Personas provide structured representations of user archetypes based on demographic, behavioural, and contextual data, enabling designers to model diverse needs and constraints in a tangible form \cite{lac2015crosscultural, car2020nudge}. Their use in health informatics has been shown to improve alignment between system functionality and real-world user practices \cite{woods2017development, haldane2019user, morton2020participatory}.

\color{darkblue}
Building on this foundation, RE research has also addressed inclusivity through human-centred, socio-technical, stakeholder-sensitive, and value-aware approaches. For example, \textit{goal-oriented RE} has been extended to incorporate stakeholders’ emotions, motivations, vulnerabilities, and non-functional concerns alongside functional goals \cite{alwidian2022towards,yu1998goal}. In parallel, \textit{value-based} and \textit{inclusive RE} approaches advocate systematically eliciting and negotiating stakeholder values, cognitive abilities, accessibility needs, and potential exclusion risks, so that requirements do not default to a narrow “average” user \cite{thew2018value,tizard2024elicitation}. Other studies have highlighted the need for RE methods that address heterogeneous and evolving requirements in complex domains such as healthcare, where users differ widely in abilities, contexts, support needs, and patterns of change over time \cite{cysneiros2002requirements}. However, this literature is less clear about how such concerns should be represented in concrete artefacts for day-to-day requirements elicitation and early analysis in chronic disease mHealth, particularly when needs shift over time and are shared across patients and caregivers.

\color{darkblue}
Complementing this RE-focused work, several \textit{persona-based inclusivity methods} originating from HCI and UX research have influenced how inclusivity concerns are operationalized in software development. The GenderMag method \cite{burnett2016gendermag} employs personas encapsulating five facets of gender differences to identify inclusivity barriers in software. Extending this idea, InclusiveMag \cite{mendez2019gendermag} generalizes the process into three steps—Scope, Derive, and Apply—enabling the creation of new inclusivity methods that address diverse dimensions such as cognitive abilities, age, vision, and socio-economic background. InclusiveMag has been used to generate inclusivity methods for smaller academic projects, including autism spectrum, age, and vision inclusivity. Similarly, SocioEconomicMag \cite{burnett2021toward,hu2021toward} was introduced to help UX professionals and developers account for variations in usage patterns associated with different socio-economic statuses.

\color{darkblue}
Together, this body of work suggests a more specific gap than our earlier framing implied. Existing RE research has already established the importance of inclusive, human-centred, and stakeholder-sensitive requirements practices, and persona-based methods have shown how inclusivity facets can be operationalized in software analysis. However, the intersection between UX/HCI-oriented inclusivity methods and RE practice needs to be articulated more explicitly. In this paper, we treat methods such as GenderMag and InclusiveMag not simply as UX evaluation techniques, but as sources of structured artefacts that can support RE activities. In particular, personas can function as requirement-relevant representations of stakeholder diversity, while walkthrough-based analyses can surface usability, accessibility, and trust-related issues that should be translated into candidate requirements, design constraints, and refinement decisions. Prior software engineering work has similarly shown that user-centred techniques such as personas, scenarios, and interaction models can contribute directly to requirements analysis and design, and that accessibility-focused RE methods can be tailored for specific user groups rather than treated only as downstream interface evaluation concerns \cite{lopes2018applying,tuunanen2010requirements}. From this perspective, ChroniUXMag contributes by adapting a UX/HCI-inspired inclusivity approach into concrete artefacts for requirements elicitation and early analysis in chronic disease mHealth, where user needs are shaped by changing health status, caregiver involvement, and long-term self-management contexts.

\rthreerev{An adaptive user interface (AUI) changes its presentation, content, or behaviour in response to user or contextual information.} \rthreeadd{In chronic-disease mHealth, this capability is relevant because user needs can vary across individuals and over time.}

\color{black}

\color{black}
\section{Method} \label{method}

\begin{figure*}
    \centering
    \includegraphics[width=1\linewidth]{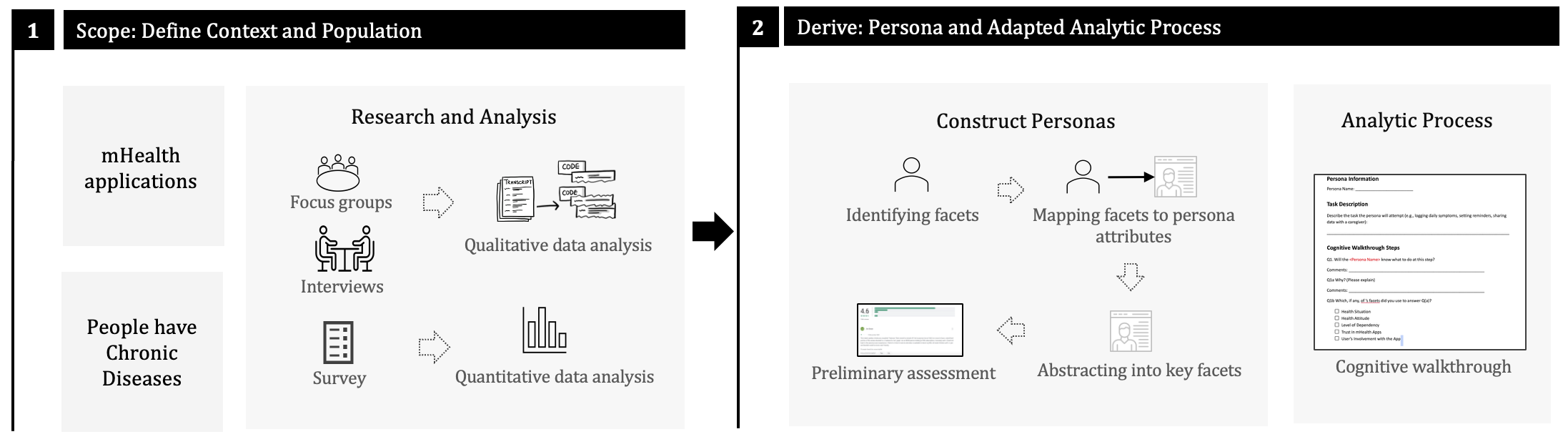}
    \caption{\rrev{Research methodology and evidence sources: systematic literature review (48 primary studies), five focus groups, seven interviews, and an online survey. The review informed scoping and instrument design.}}
    \label{fig:method}
\end{figure*}

\color{darkblue}
Our study employed a multistage empirical methodology to identify and refine user \rrev{facets} that inform inclusivity-focused requirements for mHealth applications. The study design was informed by prior persona-based inclusivity work, particularly GenderMag \cite{burnett2016gendermag} and InclusiveMag \cite{mendez2019gendermag}. Importantly, InclusiveMag is used here as a meta-method for generating a new inclusivity method, not as a standalone framework-synthesis procedure. In this study, its \textit{Scope} and \textit{Derive} stages served as conceptual guidance for identifying and structuring inclusivity-relevant factors, while the actual empirical evidence was generated through established methods: a systematic literature review, focus groups, interviews, a survey, and socio-technical grounded theory analysis. This meta-method has been used to formulate inclusive design methods such as GenderMag \cite{burnett2020doing}, AgeMag \cite{mcintosh2021evaluating}, SESMag \cite{agarwal2023socioeconomicmag}, and Elderly HealthMag \cite{xiao2026elderly}. In this sense, InclusiveMag provided the methodological scaffold, whereas the framework content was derived from empirical data and literature evidence. The Apply stage of InclusiveMag, which concerns implementation and tool-based evaluation, is left for future work (Figure~\ref{fig:method}).

\color{black}

\color{black}
\subsection{\rrev{Scope: Defining Context and Population}}

\begin{figure}[h]
    \centering
    \includegraphics[width=1\linewidth]{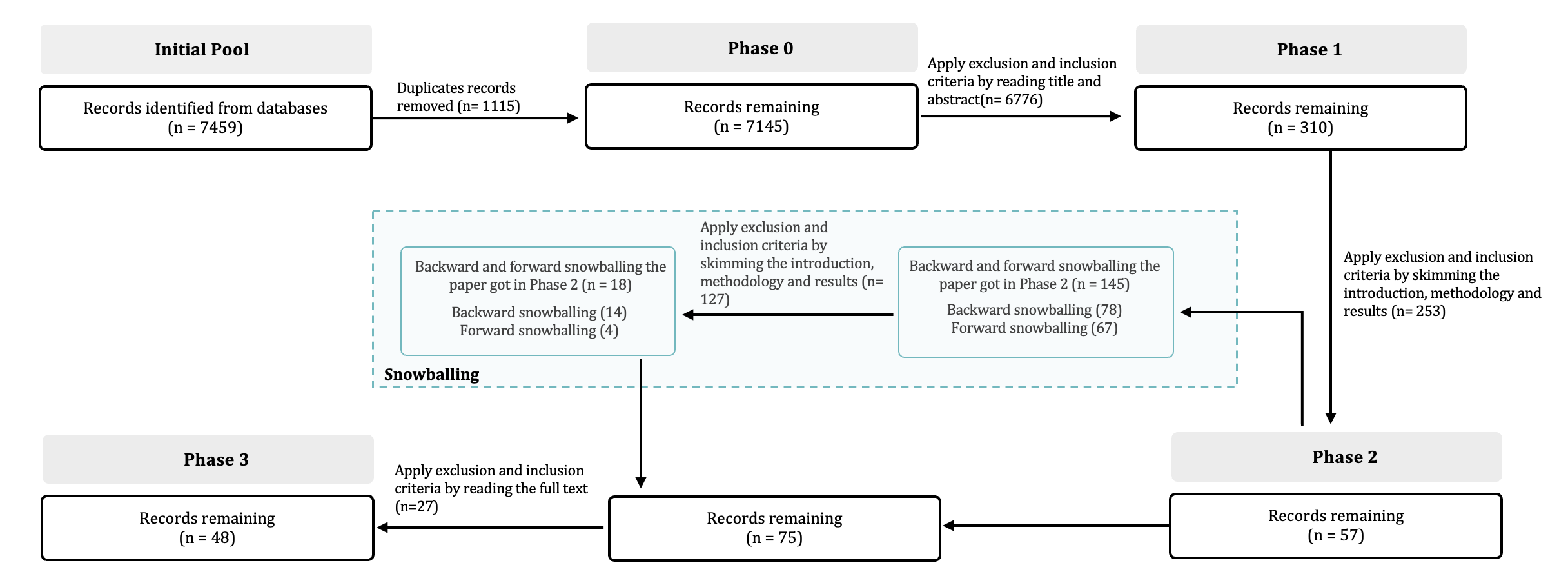}
    \caption{Stages of SLR process \cite{wang2023adaptive}}
    \label{fig:SLRIST}
\end{figure}

\color{darkblue}
We began with a Systematic Literature Review (SLR) of adaptive user interfaces in systems targeting chronic disease management \cite{wang2023adaptive}. The review followed established SLR guidelines and covered publications across software engineering, human–computer interaction, and digital health venues. Searches were conducted in major digital libraries (including ACM Digital Library, IEEE Xplore, SpringerLink,
ScienceDirect, Scopus and Medline) using combinations of terms related to adaptive user interfaces, mHealth, and chronic disease. The SLR identified and synthesized prior work on adaptation triggers, user modeling strategies, system-level adaptation mechanisms, and evaluation approaches in chronic disease–oriented systems. In total, the review included 48 primary studies after applying predefined inclusion and exclusion criteria (See Figure \ref{fig:SLRIST}). Studies were required to (1) explicitly describe an adaptation mechanism or user modeling approach, (2) target users with chronic conditions, and (3) be published in English in established venues within software engineering, human–computer interaction, or digital health. We excluded studies that focused solely on clinical decision support, medical devices without user-facing interfaces, or non-adaptive static interfaces. Papers addressing acute or short-term conditions, conceptual position papers without empirical or technical grounding, and non–peer-reviewed literature (e.g., posters, abstracts, theses) were also excluded.

\color{darkblue}

\color{darkblue}
While the review did not explicitly address accessibility or inclusivity, its findings informed the scoping of candidate user-related facets and the design of our empirical instruments. Consistent with InclusiveMag’s Scope stage, the purpose of this review was to help delimit the design space and identify candidate inclusivity-relevant factors, rather than to synthesize the ChroniUXMag framework on its own. To ground these insights in lived experiences and inclusivity concerns, we subsequently conducted qualitative and quantitative studies with individuals managing chronic conditions. In this study, inclusivity considerations emerge from empirical user data rather than directly from the SLR.

\color{darkblue}
\begin{figure}[!ht]
     \begin{subfigure}[b]{0.31\textwidth}
         \centering
         \includegraphics[width=\textwidth]{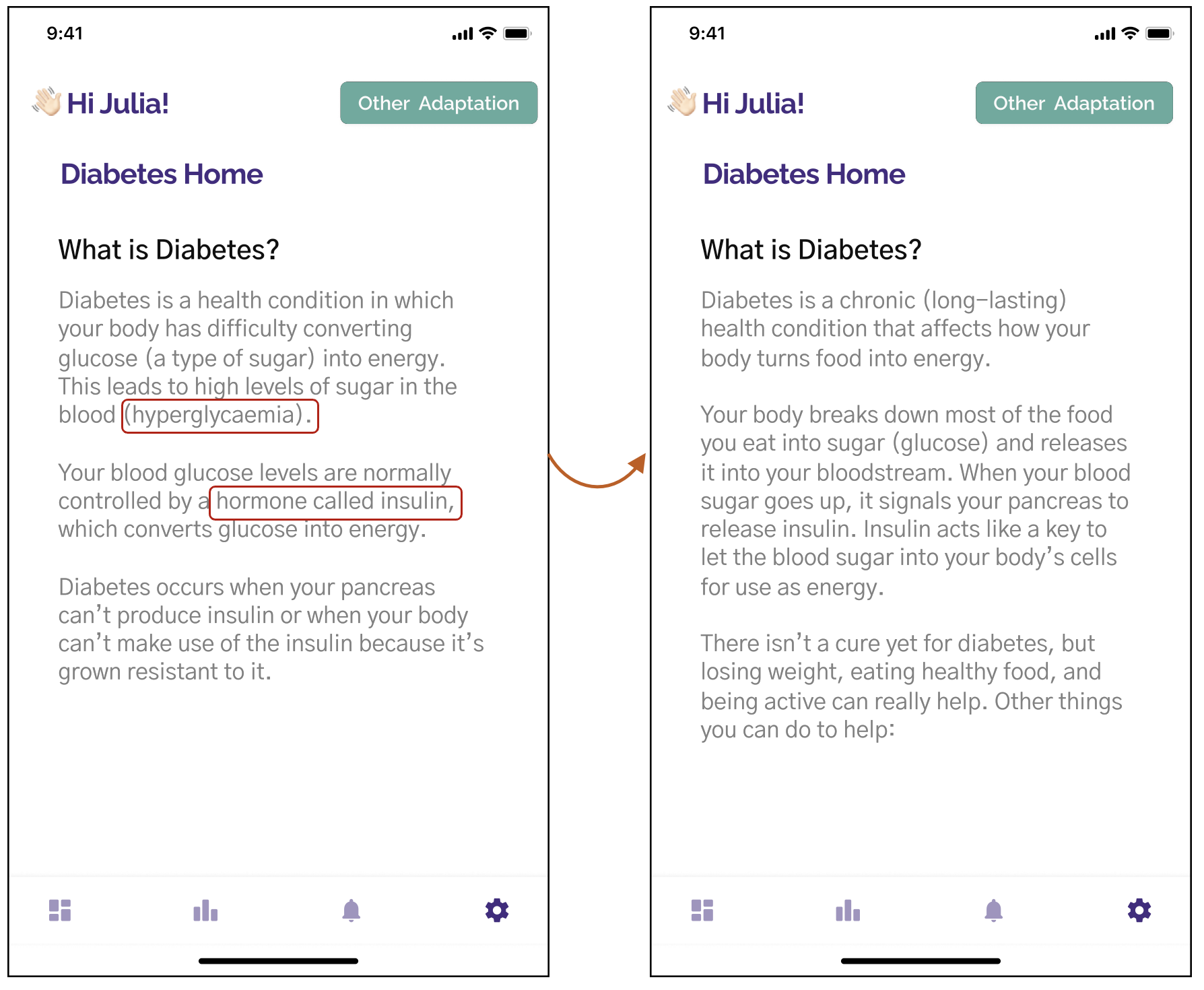}
         \caption{\scriptsize Easy-to-understand language}
         \label{(sub:preinf)}
     \end{subfigure}
     \hfill
     \begin{subfigure}[b]{0.31\textwidth}
         \centering
         \includegraphics[width=0.82\textwidth]{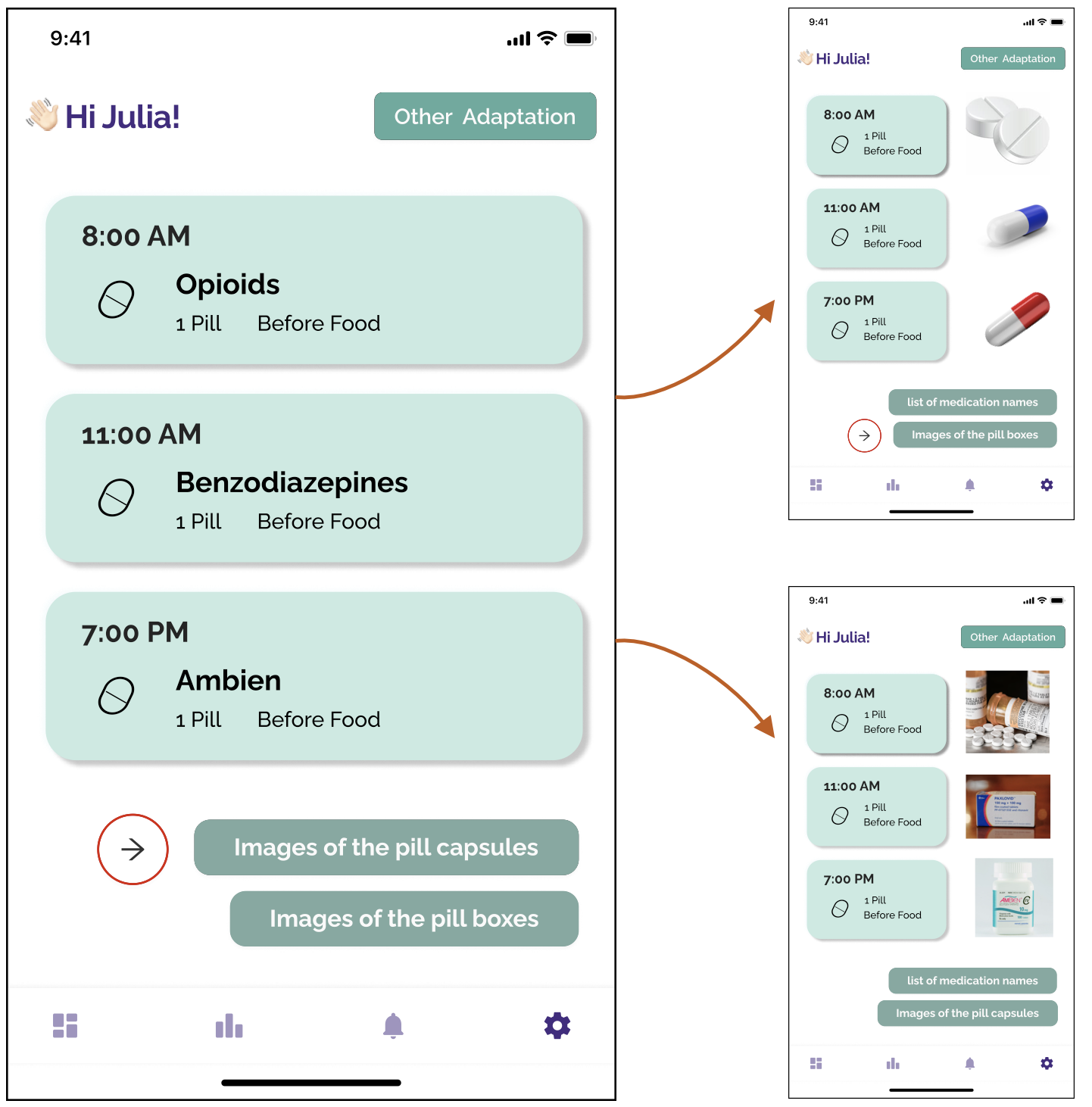}
         \caption{\scriptsize Text-to-image conversion}
         \label{(sub:conima)}
     \end{subfigure}
     \hfill
          \begin{subfigure}[b]{0.35\textwidth}
         \centering
         \includegraphics[width=0.9\textwidth]{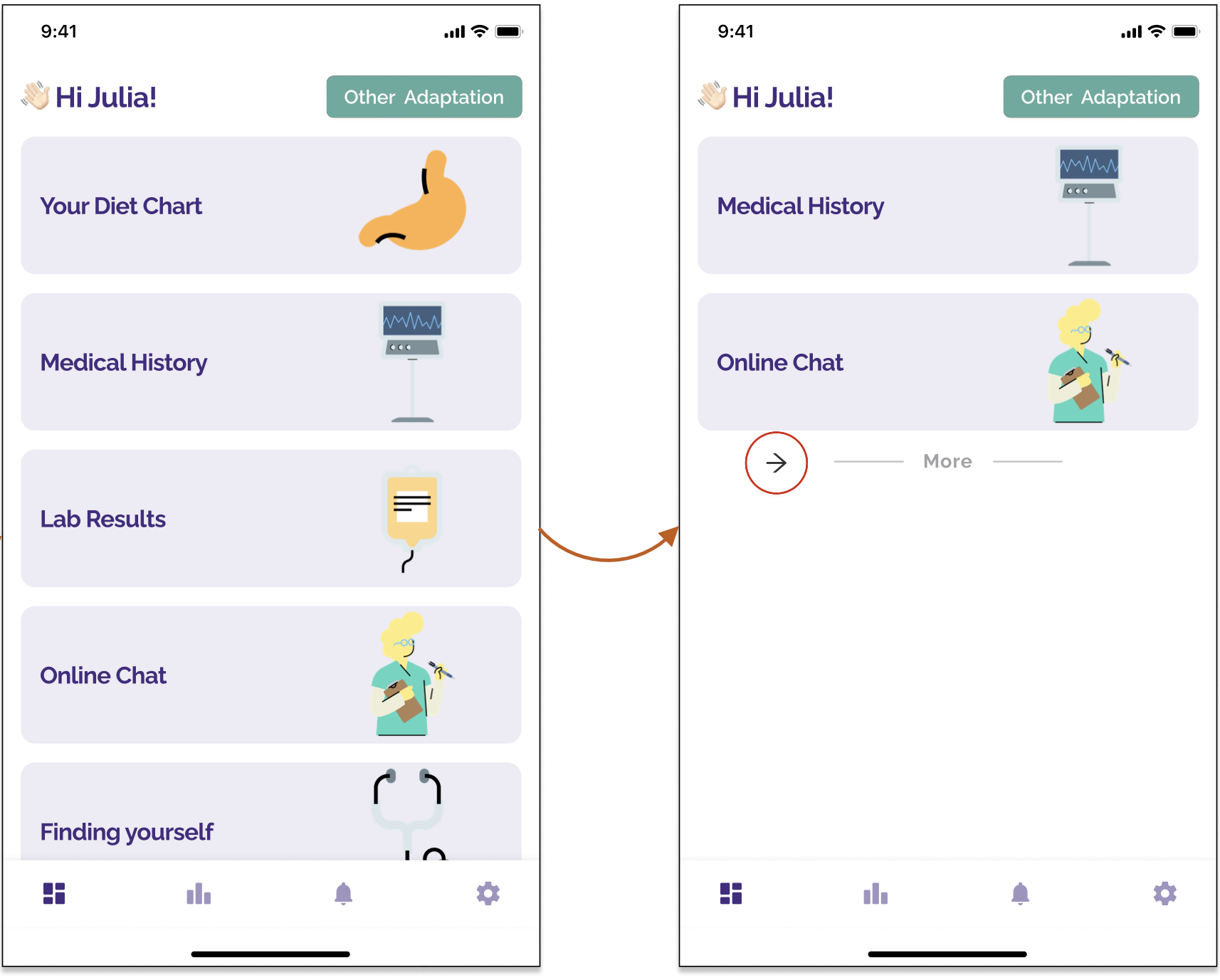}
         \caption{\scriptsize Interface elements rearrangement}
         \label{(sub:conele)}
     \end{subfigure}

\color{darkblue}
    \caption{\rrev{Examples of content adaptation in the adaptive user interface (AUI) prototype. Details of the prototype can be found in \cite{wang2024adaptivep}.}}
     \label{fig:AUIP}
\end{figure}
\rrev{\rthreerev{\textbf{Qualitative studies.} The focus groups and interviews used a common core protocol\footnote{\url{https://doi.org/10.5281/zenodo.19065215}}, adapted to group and one-to-one settings. The protocol comprised two sections.} First, we gathered demographic information about participants' chronic diseases and mHealth app use through Qualtrics. Second, participants discussed adaptations presented in an AUI prototype. The prototype demonstrated presentation adaptation (e.g., layout or visual-emphasis changes), content adaptation (e.g., tailored information), and behaviour adaptation (e.g., adaptive workflows), as identified in our systematic literature review \cite{wang2023adaptive}. Figure~\ref{fig:AUIP} shows content-adaptation examples. Participants watched a video with audio explanations and then interacted with the prototype; instructional slides and researcher support facilitated the activity.}

\color{darkblue}
\rrev{Before the main data collection, we conducted two method-specific pilots. Here, ``main data collection'' denotes the five focus groups and seven interviews whose data were included in the analysis. The focus-group pilot was conducted via Zoom with three participants who had different chronic conditions and prior mHealth experience. Two researchers served as moderator and facilitator. We used this pilot to refine prompt wording and sequencing, remove redundancy, and improve facilitation. The interviews retained the same core questions to support comparison across the two qualitative methods, but a separate pilot interview with two participants was used to refine pacing and probing for the one-to-one setting. Data from neither pilot were included in the final analysis.}

\color{darkblue}
\rrev{Theoretical sampling governed recruitment across both the focus groups and interviews \cite{hoda2021socio}. Eligibility initially required a diagnosed chronic disease and prior use of at least one self-management mHealth app. We then refined recruitment in response to emerging concepts: the first iteration emphasised older adults, the second added younger adults to contrast life-stage experiences, and the third used individual interviews to deepen and clarify the categories emerging from the focus groups. Participants were recruited through patient support groups, community health networks, online forums, and personal and professional networks.}

\radd{Chronologically, Iteration~1 comprised three focus groups with nine older participants; Iteration~2 comprised two focus groups with six younger adults; and Iteration~3 comprised seven semi-structured interviews. Thus, the qualitative sample totalled 22 participants.} \rrev{Each focus group contained three participants and lasted approximately 60--70 minutes. Two researchers attended the first two focus groups, one moderating and one facilitating;} \radd{the other three focus groups were conducted by one researcher who combined both roles.} \rrev{The interviews lasted approximately 40--60 minutes. All focus groups and interviews were conducted remotely.}

\rrev{Sampling and analysis continued until no substantially new inclusivity-relevant categories emerged. By the end of Iteration~2, the core categories from Iteration~1 recurred, with variation primarily associated with life stage. Iteration~3 provided methodological triangulation: individual interviews enabled participants to discuss personal topics without peer influence and allowed us to clarify and deepen the focus-group findings \cite{hoda2021socio,WILSON201423}.}

\color{darkblue}
\rrev{\textbf{Quantitative studies.} To examine patterns beyond the qualitative participants, we conducted an online survey\footnote{\url{https://doi.org/10.5281/zenodo.19065561}} informed by our SLR \cite{wang2023adaptive}. The expert-reviewed and pilot-tested instrument (n=5) covered demographics, health status, mHealth use, and accessibility preferences, with examples to clarify accessibility features. It was distributed through social media advertisements, hospital networks, and non-government organisations. Of 169 people who accessed the survey, 90 completed all required questions, giving a completion rate of 53\% (90/169). We used binary logistic regression to explore associations between participant characteristics and binary accessibility-preference outcomes while controlling for multiple demographic and health factors. The analysis was exploratory, not predictive or causal.}

\color{black}

\color{black}

\color{black}

\begin{figure}
    \centering
    \includegraphics[width=1\linewidth]{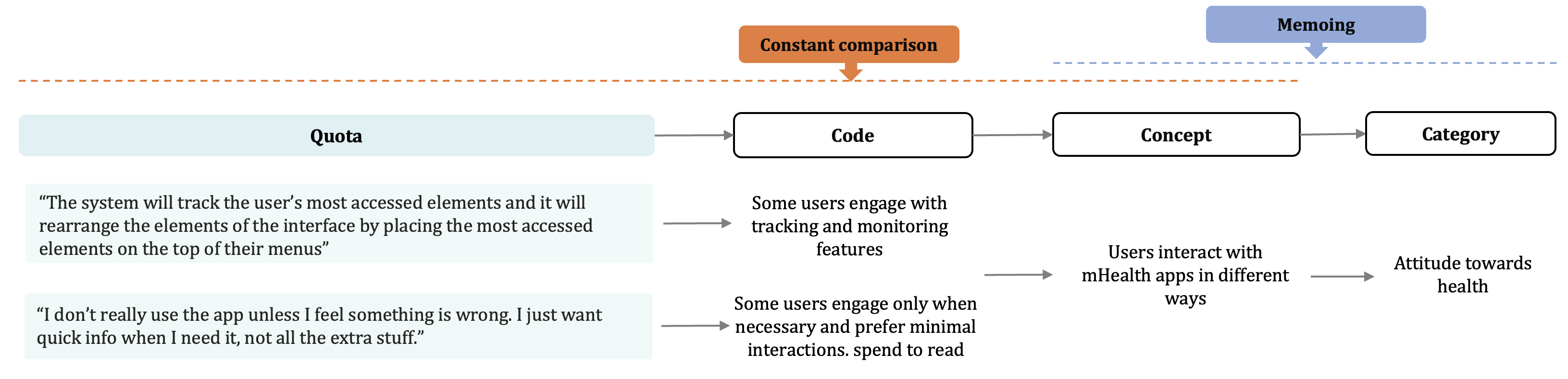}
    \caption{Examples of STGT analysis applied to qualitative data.}
    \label{fig:STGT}
\end{figure}

\color{darkblue}
\textbf{Qualitative data analysis.} We used the data analysis procedures of Socio-Technical Grounded Theory (STGT) \cite{hoda2021socio} to analyze the data from the focus group and interview study. Using STGT for data analysis, we adopted a holistic approach to data analysis, recognizing the complex and interconnected nature of the socio-technical phenomena under investigation. In this study, socio-technical phenomena refer to situations where users’ health conditions, behaviors, attitudes, and social contexts interact with technical system features such as data collection, interface adaptations, and personalization mechanisms. STGT has been applied in several studies in SE and RE to uncover socio-technical aspects, for example in understanding the role \cite{gunatilake2025role} and manifestation \cite{gunatilake2025manifestations} of empathy in SE and the influence of motivation \cite{hidellaarachchi2026understanding} and personality \cite{hidellaarachchi2024personality} in RE. For example, a user’s fluctuating health status or dependency on caregivers directly shapes how they perceive, trust, and engage with mHealth apps, while system design decisions influence users’ behaviors and expectations. STGT is therefore well suited to capturing these reciprocal relationships between human factors and technical design elements in the context of adaptive mHealth systems. The participants consented to audio transcription and the data were stored and analysed with NVivo. The first author analyzed the qualitative data and shared it with coauthors for collaborative discussion at each stage. Consistent with InclusiveMag’s Derive stage, the resulting concepts and categories were subsequently organised into higher-level facets and translated into ChroniUXMag artefacts. An example of applying STGT to data analysis is provided below and illustrated in Figure \ref{fig:STGT}.

\color{black}

\color{black}
\begin{enumerate}[left=5pt,topsep=5pt, itemsep=5pt]
\item \textbf{Open Coding and Constant Comparison:} We analysed the audio transcripts and extracted various codes from the raw data. We provide one example below.\\
    Raw Quote 1: \textit{“I always check my health data and set reminders in my app because I like to stay on top of things. It helps me keep track of my progress.”} \\ \hconcep{Code 1: some users engage with tracking and monitoring features.}\\
    Raw Quote 2:\textit{“I don’t really use the app unless I feel something is wrong. I just want quick info when I need it, not all the extra stuff.” }\\ \hconcep{Code 2: some users engage only when necessary and prefer minimal interactions.}

\color{darkblue}
\item \textbf{Constant Comparison and Concepts:} Following open coding, we applied constant comparison to examine relationships among codes both within and across data sources. Constant comparison involves continuously comparing newly identified codes with existing ones to uncover recurring patterns and meaningful distinctions in the data \cite{hoda2021socio}. For example, the two codes presented above (\hconcep{Code 1} and \hconcep{Code 2}) were repeatedly compared with each other and with related codes emerging from other transcripts. Through this process, we observed that both codes reflected differences in how users approach and engage with mHealth applications. Based on this shared pattern, we synthesized these codes into a higher-level \hconcep{\textbf{Concept:} Users interact with mHealth applications in fundamentally different ways.} We applied the same constant comparison process to the remaining codes, iteratively refining and contrasting them to identify additional concepts.

\color{darkblue}
\item \textbf{Development of Key Categories: }Drawing insights from the \textit{memos} generated during the coding process and the codes and concepts identified, the given example illustrates the \hcate{\textbf{Category:} Attitude towards health}, which captures how varying motivation levels and health awareness shape users’ expectations of, and interactions with, health systems. Memoing played a central role throughout the analysis by enabling reflective reasoning about emerging concepts and their relationships, as recommended in STGT\cite{hoda2021socio}. Memos were written iteratively during and after coding to document analytical insights, explore tentative interpretations, and trace how individual observations connected to broader socio-technical phenomena. An example memo illustrating this process is shown below.
\end{enumerate}

\color{black}
    \memo{\textbf{Memo for \hconcep{Users interact with mHealth apps in different ways.}} Participants using a proactive health strategy frequently engage with mHealth apps, utilising tracking and goal-setting features for control. Conversely, those with a reactive approach use apps mainly during symptoms or concerns, seeking quick access to essential information over customisation. These interaction styles are deeply rooted in health attitudes, which shape how users perceive and integrate mHealth applications into their daily routines. Proactive health individuals use technology for prevention and empowerment, adopting digital health solutions readily. Reactive health individuals consider mHealth tools as problem-solving aids, using them only when needed.}

\color{black}
\textbf{Quantitative data analysis.} We analysed survey data using SPSS Statistics\footnote{https://www.ibm.com/spss}, conducted chi-square tests to explore associations between demographic factors and other categorical variables, and fitted binary logistic regression models for binary accessibility-preference outcomes. Understanding how these demographic factors influence user preferences provides valuable insights into tailoring AUIs to accommodate diverse user needs and preferences. The common significance level of \begin{math}\alpha= 0.05\end{math} is chosen for statistical analysis.

\color{black}

\color{black}
\subsection{\rrev{Derive: Facets, Personas, and Adapted Analytic Process}} \label{establish dimension}

\color{darkblue}
To systematically elicit user requirements for mHealth apps that target people with chronic diseases, we referenced the \textbf{persona taxonomy} introduced by Karolita et al. \cite{karolita2023whats}, which categorizes human characteristics (known as \textit{ persona attributes} in the study) into internal and external layers. This taxonomy provides a structured framework for capturing diverse user requirements and can be adapted based on the domain of the app (e.g., healthcare, assistive technologies, education). The \rrev{facets}, derived by the first and second authors\footnote{This author has extensive experience in persona studies}, are outlined in the process described below.

\color{black}

\color{darkblue}
\begin{enumerate}[label=\textbf{\color{r2revised}\alph*.},leftmargin=*,itemsep=5pt]
\item \rrev{\textbf{Identifying first-order categories.} We applied STGT to elicit factors that influence users' experiences of mHealth adaptation. The analysis produced 13 first-order categories: \textit{mHealth user experience, mHealth app usage pattern, health conditions (multimorbidity), user involvement in the app, cultural differences, caregiver's role, digital literacy, cognitive load, accessibility needs, motivation and engagement, privacy sensitivities, age, and attitude towards health}. \radd{These categories are intermediate analytic concepts, not the final ChroniUXMag facets.}}

\color{darkblue}
\item \rrev{\textbf{Mapping categories to persona attributes.} We mapped the categories to the health-domain persona taxonomy. Conceptually related categories were grouped; for example, health conditions, cognitive load, and accessibility needs contributed to the health-situation attribute. Age and motivation/engagement remain standard fields in the persona template rather than ChroniUXMag facets \cite{karolita2024what}.}

\color{black}

\color{darkblue}
\item \rrev{\textbf{Abstracting categories into six facets.} We consolidated the first-order categories into six higher-level facets that constitute ChroniUXMag: \textit{health situation, health attitude, cultural differences, level of dependency, trust in mHealth apps}, and \textit{users' involvement with the app}. For example, health situation integrates health condition, cognitive load, and accessibility needs. Section~\ref{sec:operationalising-chroniuxmag} defines the facets and their ranges.}

\color{darkblue}
\item \rrev{\textbf{Grounding the facets in user reviews.} We used review analysis as an exploratory check of whether real-world mHealth concerns corresponded to the six facets, not as an evaluation of framework conformance. We selected four diabetes-management apps based on high ratings, substantial downloads, and free availability with optional purchases \cite{radcliffe2021pilot}. English-language reviews were pre-processed with NLTK \cite{NLTK}; facet-related keyword searches\footnote{\url{https://doi.org/10.5281/zenodo.14998546}} provided entry points for manual interpretation. We iteratively added relevant synonyms and removed ambiguous terms until no substantially new facet-relevant terms emerged.}

\color{black}
\item \rrev{\textbf{Specialising the cognitive walkthrough.} The personas are embedded in a customised evaluation protocol based on GenderMag's cognitive walkthrough (CW) \cite{burnett2016gendermag}. Evaluators simulate task steps and ask whether the persona will know what to do, recognise the correct action, and perceive progress \cite{wharton1994cognitive}. The form explicitly prompts evaluators to ground each judgement in the ChroniUXMag facets.}
\end{enumerate}

\color{darkblue}
\begin{table*}[ht!]
\centering
\caption{Participant demographic information}
\label{tab:demographics}
    \resizebox{1\textwidth}{!}{%
\begin{tabular}{p{105mm}p{100mm}}
\hline
\toprule
    \begin{tabular}[t]{p{25mm}p{70mm}}
        \multicolumn{2}{p{80mm}}{\textbf{Focus group and Interview study (n=22)}} \\ \hline
        Demographics & Details \\ \hline
        Age & 18-34 (n=7), 35-54 (n=5), 55-74 (n=10),  \\ \hline
        Gender & Females (n=15), Males (n=7) \\ \hline
        Education Level & With bachelor degree (n=17), Without bachelor degree (n=5) \\ \hline
        Nationalities & Majority being Australian (n=10), Others: Chinese, Indian, British, Filipino, Malaysian, American, and Tamil (n=12) \\ \hline
        Chronic Diseases & Type 2 diabetes (n=12), Others: Hypertension, Mental health conditions \\
    \end{tabular}
    &   \begin{tabular}[t]{p{25mm}p{70mm}}
         \multicolumn{2}{p{80mm}}{\textbf{Survey study (n=90)}} \\ \hline
        Demographics & Details \\ \hline
        Age & 18-34 (n=47), 35-54 (n=29), 55-74 (n=14) \\ \hline
        Gender & Females (n=38), Males (n=50), Prefer not to say (n=2) \\ \hline
        Education Level & With bachelor degree (n=59), Without bachelor degree(n=31) \\ \hline
        Nationalities & Majority being Australian (n=44), Chinese (n=21), USA (n=12), UK (n=6), Others:Nigeria, Canada, Korea, Spain and Sri Lanka \\ \hline
        Chronic Diseases & Cardiometabolic diseases (n=47), Immune-related (n=31),  Mental health condition (n=7), Respiratory (n=11) *\\
    \end{tabular}
    \\
        \hline  \multicolumn{2}{p{210mm}}{\textit{*These counts do not add up to 90 because some participants have multiple chronic diseases}}\\
\bottomrule
\end{tabular}}
\end{table*}

\section{\rrev{User Study Results}} \label{results}
The user study combined qualitative (focus groups and interviews, n = 22) and quantitative (survey, n = 90) data to capture diverse perspectives on mHealth use in chronic disease management. Participants varied in age, nationality, digital experience, and health background, enabling exploration of heterogeneous needs rather than representativeness of any single patient population. We acknowledge that the sample is not statistically representative and exhibits imbalances (e.g., nationality and disease types). However, the study was designed to support exploratory requirements analysis and facet identification rather than population-level generalization.

\color{darkblue}
Table \ref{tab:demographics} provides a breakdown of the characteristics of the user study participants. Interview and focus group Participants have more females (n=15) than males (n=7), ages 18-74, and various chronic diseases: type 2 diabetes (n=12), rhinitis (n=2), chronic gastritis (n=1), epilepsy (n=1), hypertension (n=2), asthma (n=3), chronic depressive disorder (n=1), mental health conditions (n=7). Ten participants have more than one type of chronic disease. All have used mHealth apps, with varying usage frequency and session durations. 17 participants hold \textit{university degrees}, with 8 bachelor's and 9 master's. Participants are numbered according to their participation in the study: \hconcep{[P1-P22]}.

\color{black}
In our survey, 56\% of the survey participants identified themselves as men. The largest age group was 25-34, accounting for 33\% of participants, followed by 21\% aged 35-44 and 16\% over 55. In terms of nationality, approximately half of the respondents identified as Australian, with Chinese and U.S. nationals accounting for 23\% and 13\% of the sample, respectively. Educationally, 44\% hold a bachelor's degree, 34\% have less than a bachelor's, and 21\% possess master's or doctorate degrees. We categorized participants' chronic health conditions into cardiometabolic, respiratory, immune-related, and mental health. Cardiometabolic diseases were the most common, affecting 52\% of participants. Notably, 24 participants reported living with multiple chronic diseases, either within the same category or across different categories, highlighting the prevalence of multimorbidity in the study population.

\color{darkblue}
\section{\rrev{Derivation of the ChroniUXMag Requirements Framework}}
Based on the qualitative analysis of data collected from focus groups and semi-structured interviews, we identified a set of categories that shape how users engage with mHealth apps for chronic disease management. We consolidate these insights into ChroniUXMag.

\subsection{\radd{Six ChroniUXMag Facets}}

\color{darkblue}
\paragraph{\rrev{Facet 1: Health Situation}} Across our empirical investigations, multimorbidity was commonly reported among participants. In the focus group and interview study, 10 out of 22 participants reported living with two or more concurrent chronic diseases, often involving combinations such as diabetes and hypertension. Similarly, in the survey study, 24 out of 90 respondents reported multiple chronic conditions. These findings highlight the prevalence of overlapping health needs among mHealth users.

\color{darkblue}
Our survey analysis further indicated that different chronic diseases were associated with variations in accessibility preferences. Examples of the accessibility features discussed in this study are illustrated in Figure \ref{fig:AUIP}. Participants with \textit{cardiometabolic diseases} showed stronger preferences for a wider range of accessibility features compared to participants with other types of chronic conditions, suggesting that their health management routines may require greater flexibility and personalisation (see Table \ref{tab:BLR}).

\begin{table}[hb!]
\centering
\renewcommand{\arraystretch}{1.1}
\caption{Binary logistic regression results of the accessibility preference and demographic aspects (\vcenteredincludep{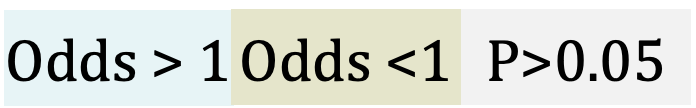})}
\label{tab:BLR}
        \resizebox{0.95\textwidth}{!}{%
\begin{tabular}
{p{30mm}p{40mm}p{35mm}p{35mm}p{35mm}p{35mm}}
\hline
\toprule
\textbf{Variables}&\textbf{Categories}&\textbf{CC*}&\textbf{AD*}&\textbf{DP*}&\textbf{MI*}\\
\hline
\multirow[c]{4}{4em}{\textbf{Nationality**}} &	UK	&	0.066/(0.16)\cellcolor{greyb}	&	0.825/(0.843)	\cellcolor{greyb}&	0.889/(0.908)	\cellcolor{greyb}&	0.133/(0.095)\cellcolor{greyb}\\
&	USA	&	0.276/(0.107)	\cellcolor{greyb}&	0.299/(0.178)	\cellcolor{greyb}&	1.504/(0.567)	\cellcolor{greyb}&	0.346/(0.193)\cellcolor{greyb}\\ &	China	&	0.125/(0.003)\cellcolor{signifl}		&	0.215/(0.025)\cellcolor{signifl}		&	1.179/(0.796)	\cellcolor{greyb}&	1.081/(0.896)	\cellcolor{greyb}\\
&	Other	&	0.783/(0.803)	\cellcolor{greyb}&	0.341/(0.264)	\cellcolor{greyb}&	1.2/(0.845)\cellcolor{greyb}&	0.533/(0.492)\cellcolor{greyb}\\
\hline
\multirow[c]{3}{4em}{\textbf{Chronic diseases}} &	Mental health	&	4.173/(0.033)\cellcolor{signif}		&	0.358/(0.093)	\cellcolor{greyb}&	2.126/(0.181)	\cellcolor{greyb}&	0.661/(0.475)	\cellcolor{greyb}\\
&Cardiometabolic&	12.215/(0.013)\cellcolor{signif}		&	8.093/(0.029)\cellcolor{signif}		&	11.264/(0.006)\cellcolor{signif}		&	1.782/(0.471)	\cellcolor{greyb}\\
&	Immune-related &	0.536/(0.554)	\cellcolor{greyb}&	3.823/(0.168)	\cellcolor{greyb}&	2.293/(0.339)	\cellcolor{greyb}&	1.463/(0.664)	\cellcolor{greyb}\\
\hline
\multicolumn{6}{p{220mm}}{{This table presents the results of binary logistic regression analyses examining how demographic factors influence user preferences for accessibility-related features in mHealth applications.

\begin{itemize}[left=0pt]
    \item Each cell shows the odds ratio (OR) for preference likelihood and the p-value for statistical significance.
    \item An OR $>1$ indicates higher preference, OR $<1$ indicates lower preference, and $p \leq 0.05$ means statistical significance.
\end{itemize}
 }}\\

\multicolumn{6}{p{220mm}}{\textit{* CC=Content complexity, AD=Add on functions, DP=Different persuasive strategy, MI=Multimodal interaction ** Australia was used as the reference category for nationality}}\\

\bottomrule
\end{tabular}}
\end{table}

\color{darkblue}
Interview and focus group data provided additional insight into how health conditions shape interaction preferences. Participants experiencing more \textit{severe} symptoms or managing \textit{multiple} chronic diseases often preferred simplicity and greater system assistance, highlighting the importance of predictable and stable interface behaviour. One participant explained, \hquote{“I don’t want things changing all the time—I already have enough unpredictability with my health."}[P4] Similarly, users with advanced age and complex health conditions preferred adaptations that were limited to specific, easy-to-understand parts of the interface to reduce cognitive load [P9, P19, P20].

\color{darkblue}
In contrast, participants with \textit{milder} or more \textit{stable} conditions tended to value flexibility and personalisation. For example, one participant noted \hquote{“I’d rather have the app adjust to my routine rather than forcing me to use features I don’t need.”}[P7] Another participant experiencing \textit{fluctuating} symptoms highlighted the need for dynamic system behaviour, stating \hquote{“Some days, I need more reminders and tracking, but on good days, I don’t want the app to overwhelm me with notifications.”}[P9]

\color{darkblue}
To interpret these patterns, we draw on prior research suggesting that chronic disease management often involves fluctuating symptoms and varying levels of daily functioning \cite{di2019chronic, paymal2024good}. Such variability can influence how users interact with digital health technologies and the level of assistance they expect from them. Users managing more complex or unpredictable health situations may prioritise stability, simplicity, and system guidance, whereas users with more stable conditions may prefer greater flexibility and personal control. These findings suggest that diverse health situations significantly shape the \textbf{goals} users seek to achieve with mHealth applications and highlight the need for adaptable interface designs that can accommodate changing health conditions over time.

\color{darkblue}
\paragraph{\rrev{Facet 2: Health Attitude}} Our survey results indicate that participants with higher levels of education had significantly higher odds of preferring motivational data use in mHealth applications (Bachelor: OR = 7.835, $p = 0.006$; Postgraduate: OR = 6.798, $p = 0.035$; Table \ref{tab:BLRData}). Interview data further suggest that these participants often viewed adaptive features and personalised insights as helpful tools for supporting their health goals. For example, one participant emphasised the value of feedback derived from behavioural patterns, stating \hquote{“I like when the app gives me insights based on my habits—it helps me stay on track with my health goals.”}[P10] Another participant highlighted how adaptive features can reinforce engagement, explaining \hquote{“If the app adjusts to my progress and suggests new challenges, it keeps me engaged and pushes me to do better.”}[P20]

\color{darkblue}
In contrast, some participants demonstrated a more reactive attitude towards health management. These users tended to interact with mHealth applications primarily in response to symptoms or immediate concerns rather than as tools for continuous self-management. For instance, one participant expressed scepticism about the value of adaptation features, stating \hquote{“The adaptation itself doesn’t automatically make the app more useful. It is really about whether it matches my own motivation to use the app. Just trying to make me like the app more or use it more doesn’t really help.”}[P4] Participants with this reactive attitude often reported tolerating certain system limitations as long as the app provided essential support when needed.

\color{darkblue}
To interpret these patterns, we draw on the concept of \textit{health attitude}, which refers to individuals’ awareness of and willingness to actively engage in behaviours that support their health and well-being \cite{dutta-Bergman2004healthattitude, behnaz2024factors}. Prior research suggests that individuals with higher levels of education often demonstrate greater health literacy and awareness of the potential benefits of digital health technologies \cite{woldaregay2018motivational}. Such users may therefore be more receptive to personalised recommendations and adaptive feedback, perceiving these features as useful tools for supporting long-term health management. Conversely, individuals with a more reactive health attitude may engage with mHealth applications primarily when problems arise, which can reduce their perceived value of proactive features such as behavioural insights or adaptive motivation mechanisms \cite{yan2021mobileapps}. As previous studies have shown, higher levels of health consciousness are associated with greater adoption of mHealth applications and stronger intentions to continue using them \cite{jacob2022sociotechnical}.

\color{black}

\color{darkblue}

\begin{table}[h]
\centering
\renewcommand{\arraystretch}{1.3}
\caption{Binary logistic regression results of the data types and demographic aspects (\vcenteredincludep{Figure/Odds.png})}
\label{tab:BLRData}
        \resizebox{0.95\textwidth}{!}{%
\begin{tabular}{p{39mm}p{30mm}p{30mm}p{30mm}p{30mm}p{30mm}p{30mm}p{30mm}}
\hline
\toprule
\textbf{Variable}&\textbf{Category}&\textbf{Physiological characteristics}&\textbf{Physical characteristics}&\textbf{Preference}&\textbf{Feedback}&\textbf{Goals}&\textbf{Motivation}\\
\hline
\multirow[c]{4}{4em}{\textbf{Nationality}} &	UK	&	0.093/(0.053)\cellcolor{greyb}&	0.042/(0.019)\cellcolor{signifl}	&	0.047/(0.018)\cellcolor{signifl}	&	1.618/(0.633)	\cellcolor{greyb}&0.222/(0.899)	\cellcolor{greyb}&	0.727/(0.767)	\cellcolor{greyb}\\
&	USA	&	0.121/(0.009)\cellcolor{signifl}	&	0.032/(0.002)\cellcolor{signifl}	&	0.138/(0.018)\cellcolor{signifl}	&	0.205/(0.07)	\cellcolor{greyb}&0.11/(0.015)\cellcolor{signifl}	&	0.168/(0.052)	\cellcolor{greyb}\\
&	China	&	0.303/(0.066)	\cellcolor{greyb}&	0.485/(0.283)	\cellcolor{greyb}&	0.127/(0.003)\cellcolor{signifl}	&	0.216/(0.016)\cellcolor{signifl}	&	0.129/(0.005)	\cellcolor{signifl}&	0.048/($<0.001$)	\cellcolor{signifl}\\
&	other	&	0.224/(0.092)	\cellcolor{greyb}&	0.736/(0.747)	\cellcolor{greyb}&	0.624/(0.647)	\cellcolor{greyb}&	0.565/(0.512)	\cellcolor{greyb}&	0.238/(0.122)	\cellcolor{greyb}&	0.101/(0.057)\cellcolor{greyb}	\\
\hline
\multirow[c]{2}{5em}{\textbf{Education level}} &	Bachelor	&	1.023/(0.97)	\cellcolor{greyb}&	1.09/(0.896)	\cellcolor{greyb}&	0.952/(0.937)	\cellcolor{greyb}&	1.428/(0.544)	\cellcolor{greyb}&	1.1/(0.881)	\cellcolor{greyb}&	7.835/(0.006)	\cellcolor{signif}\\
&	Postgraduate	&	2.116/(0.355)	\cellcolor{greyb}&	1.568/(0.602)	\cellcolor{greyb}&	1.71/(0.513)	\cellcolor{greyb}&	1.511/(0.569)	\cellcolor{greyb}&	1.634/(0.52)	\cellcolor{greyb}&	6.798/(0.035)\cellcolor{signif}	\\
\hline
\multicolumn{8}{p{230mm}}{This table presents the results of binary logistic regression analyses examining how demographic factors influence preferences for how mHealth applications handle and make use of personal health data.

\begin{itemize}[left=0pt]
    \item Each cell shows the odds ratio (OR) for preference likelihood and the p-value for statistical significance.
    \item An OR $>1$ indicates higher preference, OR $<1$ indicates lower preference, and $p \leq 0.05$ means statistical significance.
\end{itemize}}\\

\multicolumn{8}{p{170mm}}{\textit{* Odds Ratios/(P-value)}}\\
\bottomrule
\end{tabular}}
\end{table}

\paragraph{\rrev{Facet 3: Cultural Differences}} The analysis provided preliminary evidence that cultural background may be associated with variations in user preferences for accessibility-related features in mHealth applications. For example, relative to the Australian reference group used in the regression model, participants with Chinese nationality showed significantly lower odds of preferring higher content complexity (OR = 0.125, $p = 0.003$) and add-on functions (OR = 0.215, $p = 0.025$), suggesting that cultural background may influence perceptions of how health information should be presented (Table \ref{tab:BLR}). Participants from the UK, USA, and China also showed differing patterns regarding how applications use personal data, including physiological attributes, goals, and motivation (Table \ref{tab:BLRData}). However, the small \textit{sample size} of UK and USA participants and the exploratory nature of the survey analysis mean that these observations should be interpreted as preliminary rather than conclusive (see Table \ref{tab:demographics}).

\color{darkblue}
To interpret these differences, we draw on Hofstede’s cultural dimensions \cite{hofstede2001culture}, which provide a widely used framework for understanding cross-cultural variation in values and interaction preferences. Cultural context has been shown to influence how users perceive, accept, or resist digital systems \cite{alsswey2020culture}. In the context of mHealth applications for chronic disease management, these cultural differences may shape expectations about how health information is presented, how decisions are supported, and how personal data are used. Two cultural dimensions are particularly relevant for interpreting the observed patterns:

\color{darkblue}
{\color{r2revised}\begin{itemize}[leftmargin=*]
    \item \textit{Individualism vs. Collectivism}: This dimension reflects whether a society prioritises individual autonomy or collective decision-making \cite{hofstede2001culture}. In individualistic cultures, users may prefer greater personal control over health information and app configurations, including personalised feedback and independent data management. In contrast, users from more collectivist cultural contexts may place greater value on features that facilitate shared health management, such as communication with family members or integration with healthcare providers. These differences may help explain variations observed in participants’ preferences for how mHealth apps use personal data and support health decision-making.

\color{r2revised}
    \item \textit{Long-Term vs. Short-Term Orientation}: This dimension describes whether a culture tends to emphasise long-term planning and sustained effort or prioritises immediate outcomes \cite{hofstede2001culture}. In the context of chronic disease management, long-term-oriented users may prefer features that support gradual behavioural change, long-term tracking, and predictive health insights. Conversely, users with a stronger short-term orientation may prioritise immediate feedback, simplified summaries, or short-term goal tracking \cite{guthrie2009time}. These differences may partly explain variations in preferences regarding content complexity and how health information is structured within the application.
\end{itemize}}

\color{darkblue}
\paragraph{\rrev{Facet 4: Level of Dependency}} Participants frequently described different levels of reliance on mHealth applications when managing their chronic conditions. Some participants demonstrated a \textit{high level of dependency} on these apps, relying on them as essential tools for daily health management. In such situations, applications were expected to provide simplified functional adaptations that reduce cognitive load and present clear and direct insights. Caregivers often played a central role in configuring and personalising the app to meet the patient’s needs. As one participant explained: \hquote{“My husband has dementia... deciding what features he needs is a bit tough for him. Adding more stuff for him to remember is just not doable. I might need to help my husband set up things to suit his needs because I know what he needs better than he does."} [P16]

\color{darkblue}
Caregivers, who are familiar with patients’ needs and limitations, frequently manage application settings on behalf of patients. However, incorporating family members into this process also introduced additional complexity, as highlighted in the interview and focus group data. Several participants described situations where they attempted to understand the accessibility features configured for their parents, which sometimes diverged from the parents’ own priorities [P8--P10, P16, P18--P20, P22]. For instance, one participant responsible for remotely monitoring a technologically inexperienced elderly mother highlighted potential risks, such as \hquote{“misinterpreting data or making incorrect health decisions."} [P20]. Participants also described interactions with \textbf{healthcare providers}, which created an additional layer of dependency on medical expertise and specialised terminology. One participant noted this challenge, stating: \hquote{“I often talk to doctors, and they use medical terminology. Understanding it is important to have a meaningful discussion, but it can also be overwhelming. At the same time, I’m surrounded by people who rely on complex medical phrases, so I have to find a balance.”} [P18] Some mHealth applications, such as Health2Sync\footnote{https://www.health2sync.com/} and LibreLinkUp\footnote{https://www.librelinkup.com/}, also incorporate caregivers directly into the chronic disease management process.

\color{darkblue}
At the same time, participants highlighted tensions between reliance on caregivers and clinicians and concerns about \textbf{privacy and data sharing}. For example, one participant emphasised that they would only share health data if they clearly understood how the data would be used and retained the ability to stop sharing it when needed [P17]. In contrast, other participants demonstrated a \textit{lower level of dependency} on mHealth applications. These users described the apps primarily as tools for independent health monitoring and self-management. They often sought features that support autonomy, flexibility, and deeper personal insight. Several participants expressed a desire to better understand their health data by \textit{integrating multiple data sources}, which could help them derive insights and manage their health goals more effectively [P12--P15, P20--P22].

\color{darkblue}
These observations can be interpreted through prior work that distinguishes different levels of dependency among individuals managing chronic conditions \cite{schiavo2022trade}. In ChroniUXMag, dependency refers to the extent to which a user’s effective engagement with an mHealth application relies on support from others or on app-mediated scaffolding. In our data, this facet spans three closely related sub-dimensions: \textit{functional dependence} (reliance on simplified interactions or system guidance), \textit{caregiver-mediated dependence} (reliance on family members or carers to configure, interpret, or use the app), and \textit{clinician-mediated dependence} (reliance on healthcare professionals to explain terminology, interpret data, or support decision making). Users with higher dependency may require simplified interactions and greater caregiver involvement, while users with lower dependency may prioritise autonomy and self-management capabilities. Furthermore, the \textit{\textbf{progressive nature of chronic diseases}} can alter these relationships over time \cite{lorig2003self}. As individuals gradually experience changes in independence, their reliance on mHealth applications and accessibility features may evolve. This progression highlights the importance of designing mHealth applications with flexibility and scalability so that they can accommodate changing levels of dependency throughout the course of disease management.

\color{darkblue}
\paragraph{\rrev{Facet 5: Trust in mHealth Apps}} Participants frequently discussed trust when reflecting on their experiences with mHealth apps and their accessibility features. In particular, trust appeared closely related to users’ familiarity with digital technologies and their expectations of how effectively these apps could support their needs. Participants who reported greater familiarity with smartphones or health apps tended to express more confidence in exploring and adjusting accessibility features. In contrast, less experienced users often described uncertainty and hesitation when interacting with adaptive functions. As one participant explained, \hquote{“I don’t know how these changes happen, and that makes me uncomfortable. I’d rather keep things the way they are.”} [P19]. Several participants also indicated that they were comfortable with the app making adjustments, but only if they understood why the changes were occurring [P17, P19, P20].

\color{darkblue}
Interpreting these findings through the lens of trust helps explain these patterns. In this study, we adopt Komiak and Benbasat’s definition of trust as the rational belief that a trustee can be relied upon \cite{komiak2006effect}. Trust plays an important role in shaping users’ perceptions of technology and their willingness to adopt or engage with digital systems \cite{jarvenpaa1999consumer}. Our findings suggest that familiarity with technology may influence how users develop such trust in mHealth apps. Prior research indicates that experience with digital systems can shape users’ expectations of system behaviour and increase their confidence in interacting with technological features \cite{venkatesh2003user}. Familiarity may also reduce cognitive load when navigating interface options, making experienced users more comfortable experimenting with accessibility settings. Conversely, limited familiarity can create uncertainty about how features operate, which may weaken trust and discourage exploration.

\color{darkblue}
Our findings further suggest that transparency and user control are important mechanisms for fostering trust in adaptive mHealth systems. Participants emphasised the need to understand why the system was making changes, indicating that transparent explanations can reassure users that the system operates in their best interests. Similarly, user control influenced trust, as some individuals preferred the ability to review, override, or customise system behaviour rather than relying entirely on automated processes. Prior research also highlights the importance of both cognitive and emotional trust in mobile technology adoption  \cite{idemudia2014influence}. In the context of mHealth, trust may be particularly critical for vulnerable groups, such as older adults or individuals managing chronic conditions, who rely on these systems for ongoing health support \cite{meng2019investigating, koo2023factors}. By incorporating clear explanations, flexible customisation options, and inclusive design practices, mHealth apps can better support trust while accommodating diverse levels of digital experience and accessibility needs.

\color{darkblue}
Our findings further suggest that \textit{transparency} and \textit{user control} are important mechanisms for supporting inclusive interaction with mHealth systems by fostering user trust. Participants frequently emphasised the need to understand why the system was making changes, indicating that transparent explanations help users make sense of adaptive behaviour and reduce uncertainty. Such transparency is particularly important for users with varying levels of digital literacy or health knowledge, as it allows them to better interpret system actions and decide whether they are appropriate for their needs. Similarly, \textit{user control} was closely related to trust. Several participants expressed a preference for being able to review, override, or customise system behaviour rather than relying entirely on automated processes. Providing such control enables users with different health conditions, levels of technological experience, or personal preferences to tailor the system to their own circumstances.

\color{darkblue}
Prior research also highlights that both cognitive and emotional trust play an important role in mobile technology adoption \cite{idemudia2014influence}. In the context of mHealth applications, trust may be particularly critical for vulnerable populations, such as older adults or individuals managing chronic conditions, who often depend on these systems for ongoing health support \cite{meng2019investigating, koo2023factors}. By incorporating clear explanations of system behaviour and flexible options for user control, mHealth applications can better support inclusive use while accommodating diverse levels of digital experience, health literacy, and accessibility needs.

\color{darkblue}
\paragraph{\rrev{Facet 6: User's Involvement with the App}} In our interview and focus group study, we identified two distinct modes of engagement with mHealth apps. Some participants demonstrated \textit{passive involvement} with the app. These participants reported limited interaction with advanced features and little interest in exploring the system’s capabilities. For example, some admitted that they \hquote{“have not fully explored what the app can do”,} while others expressed \hquote{“a lack of concern about the system's adaptation process”.} Participants in this group often indicated tolerance of automated behaviours and a preference for minimal interaction with the app [P1--P3, P5, P9, P17, P19, P21]. In contrast, other participants showed \textit{active involvement} with the app. These users described experimenting with different application features and adjusting data inputs to better understand how the system responds. Their engagement was often motivated by a desire for greater transparency, personalisation, and control over the app’s behaviour [P4, P6, P7, P8, P11, P18].

\color{darkblue}
These observations can be interpreted through prior research on user involvement in intelligent systems. User involvement has been shown to influence the acceptance of intelligent technologies, as individuals who actively engage with system features are more likely to develop an understanding of how the system operates and how it supports their goals \cite{eiband2021support}. In the context of mHealth apps, varying levels of involvement may therefore shape how users perceive adaptive features, the degree of control they expect, and their willingness to rely on automated system behaviours.

\color{black}
\subsection{\rrev{Exploratory Grounding of Key Facets in User Reviews}}
\rrev{Four apps were selected for exploratory grounding of the ChroniUXMag requirements framework. Details about the chosen apps are provided in Table \ref{tab:diabetes_apps}. A total of 2071 reviews were categorised to reflect the relevance of the proposed facets. We partially adopt the approach by Sj{\o}berg et al. \cite{sjoberg2008building} to evaluate the theoretical foundations used to perform the preliminary assessment against six facets of ChroniUXMag. Specifically, we applied the \textbf{parsimony} criterion, which refers to the degree to which a theory is economically constructed with a minimal number of concepts and propositions.}

\color{black}
\begin{table}[h]
    \centering
    \caption{Tagged reviews for different applications}
        \resizebox{0.9\textwidth}{!}{%
    \begin{tabular}{p{20mm}p{90mm}p{40mm}p{30mm}}
    \hline
    \textbf{\#} &\textbf{App Name} & \textbf{Rate/ Downloads} &\textbf{Tagged review} \\ \hline
    \textbf{App 1} &mySugr - Diabetes Tracker Log\textsuperscript{1} & 4.6/5M+ & 1753\\
    \textbf{App 2} &Gluroo: Diabetes Log Tracker \textsuperscript{2} &4.3/50k+& 124\\
   \textbf{App 3} & Health2Sync - Diabetes Tracker\textsuperscript{3} &4.6/1M+& 162\\
   \textbf{App 4} & LibreLinkUp\textsuperscript{4} &4.6/1M+& 32 \\\hline
     & && 2071\footnotesize(Sum)\\
   \hline
 \multicolumn{4}{p{165mm}}{\scriptsize \textit{1.https://www.mysugr.com/en/diabetes-app 2.https://gluroo.com/ 3. https://www.health2sync.com/4.https://www.librelinkup.com/}}\\
\hline
    \end{tabular}}

    \label{tab:diabetes_apps}
\end{table}

\color{black}
\rrev{To support this grounding, we filter user reviews using keywords corresponding to each facet, resulting in reviews being categorised into six groups based on the ChroniUXMag facets. We then manually analysed the keywords to determine whether all six facets were essential, in accordance with the\textit{ parsimony criterion}. Our analysis revealed that all facets were presented and discussed in user reviews, providing preliminary evidence of their relevance, and the reviews also reflected the range of each facet. A summary of the reviews for each facet can be found in Table \ref{tab:dimensionsnum}.}

\color{black}
\paragraph{\rrev{Facet 1: Health Situation}}
Within the health situation \rrev{facet}, users frequently mention how their symptoms and disease management needs impact their interaction with mHealth apps (18\% of tagged reviews). The health situation reflected in the user reviews \textit{varies widely}, ranging from early diagnosis to more advanced stages with multiple complications. For instance, one review stated, \hquote{“This is an easy, and almost fun, way to track my blood sugar. It's even a way to track meals and insulin injections,"} suggesting that the user may be in the \underline{middle stage} of diabetes and appreciates the app’s ability to track diet and medication.

\color{black}
\paragraph{\rrev{Facet 2: Health Attitude}}
Users in the tagged reviews generally exhibit a \textit{positive health attitude}, likely because those who leave reviews tend to be frequent users who are actively engaged with their health management (28\% of reviews). One review mentioned \hquote{“This app is great, ideal for logging levels quickly instead of a book and pen every meal time. The on-screen info is a fantastic help. Makes me competitive against myself to beat my own levels and improve them all the time."} This suggests that the user has a \underline{high level of health awareness}, is actively engaged with the app, and uses it diligently to track and improve their health metrics.

\color{black}
\paragraph{\rrev{Facet 3: Cultural Differences}}
Culturally related comments are relatively \textit{rare} in the reviews, accounting for only 7\% of all tagged reviews. Focusing on \textit{English}-language reviews might limit insights from culturally influenced health preferences, but valuable patterns still emerge. One review stated, \hquote{“I'm a newly diagnosed diabetic, so it's kinda like having a personal cheerleader. It is nice to have all your info at the palm of your hands."} This indicates that the user may have a \underline{collectivist} mindset, valuing community support and encouragement throughout their health journey.

\color{black}
\paragraph{\rrev{Facet 4: Level of Dependency}}
Dependency is a common theme in user reviews, with 32\% of tagged reviews discussing how mHealth apps help users manage the health of their \textit{partners, parents, or children} and aiding communication with \textit{healthcare professionals}. One review stated, \hquote{“My 9 year old daughter is a T1D. She loves this app...Easily sent as a pdf to our Drs at any time! ...I find the charts and stats extremely helpful especially hypo/ hyper average. DR said it would be better for them if highs and lows were different colors instead of both being red...!".} This suggests that the user is a caregiver for their child and also the primary user of the app, indicating that the child is \underline{fully dependent}, while the parent utilises the app to monitor and share health status with family members and healthcare providers.

\color{black}
\paragraph{\rrev{Facet 5: Trust in mHealth Apps}}
For the trust \rrev{facet}, users are primarily worried about \textit{privacy}, especially the reasons and usage of data collection, which make up 36\% of review comments. One review stated, \hquote{“I don't trust apps that want any personal information."} This suggests that the user has \underline{low trust} in the app, because it requests personal information.

\color{black}
\paragraph{\rrev{Facet 6: User's Involvement with the App}}
Users show a very positive involvement into use of the \textit{customisation} functionalities, as reflected in 34\% of user reviews. Many reviewers  actively explore the app’s customisation options and provide both positive and negative feedback on its flexibility. For instance, one user praised the customisation aspect, stating, \hquote{“I love that I can personalise the app, and the little green monster makes tracking fun—it feels interactive and engaging!”} Meanwhile, another user expressed frustration, saying, \hquote{“I wish I could modify or remove certain fields to better fit my needs, but the app doesn’t allow it.”} This all indicates that the user has \underline{active involvement with the app}, demonstrating a willingness to explore its features and engage with its functionality.

\color{black}
\begin{table}[h]
    \centering
    \caption{\rrev{Summary of ChroniUXMag Facets in User Reviews}}
    \label{tab:dimensionsnum}
            \resizebox{0.9\textwidth}{!}{%
    \begin{tabular}{p{50mm}p{10mm} ll}
        \hline
        \textbf{\rrev{Facet}} & \textbf{\#} & \textbf{\% of reviews \textsuperscript{1}} &\textbf{Example Keywords/Phrases} \textsuperscript{2}\\
        \hline
        Health situation & 372 & \mybarhhigh{5.31}{18\%} &\textit{"symptom", "sugar spike", "fatigue"}\\
        Health attitude & 583 & \mybarhhigh{8.33}{28\%} &\textit{"aware", "behaviour", "engage"}\\
        Cultural differences & 136 & \mybarhhigh{1.94}{7\%} &\textit{"community", "personal", "language"}\\
        Level of dependency & 667 & \mybarhhigh{9.53}{32\%} &\textit{"caregiver","son", "daughter"}\\
        Trust in mHealth apps & 749 & \mybarhhigh{10.70}{36\%} &\textit{"privacy", "concern", "cautious"}\\
        User’s involvement with the apps. & 700 & \mybarhhigh{10.00}{34\%} &\textit{"setup", "automatic", "customise"}\\
        \hline
        Total reviews & 2071 & &\\
        \hline
        \multicolumn{4}{p{170mm}}{\scriptsize \textit{1. Percentages represent facet occurrences relative to 2071 tagged reviews, rather than distinct review counts. The total row gives the review corpus size.
   }}\\
           \multicolumn{4}{p{170mm}}{\scriptsize \textit{2. The full list of keywords is available in the supplementary materials: https://doi.org/10.5281/zenodo.14998546
   }}\\
        \hline
    \end{tabular}}

\end{table}

\color{darkblue}
The alignment between recurring themes in user reviews and the ChroniUXMag facets indicates that these facets capture salient user concerns expressed in real-world mHealth usage. Rather than demonstrating the structural quality of the framework itself, this analysis provides preliminary empirical support for the relevance and coverage of the identified facets as a basis for reasoning about user adaptation needs in mHealth applications.

\color{darkblue}
\section{ChroniUXMag Framework and Operationalisation} \label{sec:operationalising-chroniuxmag}
To make the framework explicit, Table~\ref{Table: dimensions legend} summarises the ChroniUXMag facets and the facet ranges used in our analysis and artefact construction.
{
\tiny
\setlength{\tabcolsep}{4pt}
\renewcommand{\arraystretch}{1.15}
\setlength\LTcapwidth{\linewidth}   %
\setlength\LTleft{0pt}\setlength\LTright{0pt} %

\newcolumntype{P}[1]{>{\RaggedRight\arraybackslash}p{#1}}

\begin{longtable}{
  P{\dimexpr.22\linewidth\relax}%
  P{\dimexpr.18\linewidth\relax}%
  P{\dimexpr.30\linewidth\relax}%
  P{\dimexpr.30\linewidth\relax}%
}
\caption{ChroniUXMag legend}
\label{Table: dimensions legend}\\
\toprule
\textbf{\rrev{Facets}} & \textbf{Range} & \textbf{About} & \textbf{Example} \\
\midrule
\endfirsthead

\multicolumn{4}{l}{\textit{Table \ref{Table: dimensions legend} (continued)}}\\
\toprule
\textbf{\rrev{Facets}} & \textbf{Range} & \textbf{About} & \textbf{Example} \\
\midrule
\endhead

\midrule \multicolumn{4}{r}{\textit{Continued on next page}}\\
\midrule
\endfoot

\bottomrule
\endlastfoot

\textbf{Health situation} & Early stage &
Users experience mild or barely noticeable symptoms, occasional discomfort, and minimal disruption in their daily routine. &
\textit{Middle stage}: Daniel is managing high blood pressure, but his reluctance to follow medical advice has led to complications, including the onset of kidney issues. \\

 & Middle stage &
Users experience progressive symptoms which cause limitations in daily activities and require regular medical intervention or support. & \\

 & Severe stage &
Users experience complications (multimorbidity), substantial difficulty in carrying out daily tasks, and often require full-time care or support. & \\

\addlinespace

\textbf{Health attitude} & Low awareness &
Users demonstrate limited interest in health management, tend to avoid tasks related to health, possess a limited understanding of their condition, lack motivation to make changes, and engage only when symptoms become unmanageable. &
\textit{Moderate awareness}: Daniel is aware of his health challenges but struggles to follow his doctor’s recommendations, worsening his hypertension. \\

 & Moderate awareness &
Users demonstrate a moderate interest in managing their health, sporadically participate in health-related activities, and have a basic grasp of their condition but may not routinely act on it or prioritise preventive measures. & \\

 & High awareness &
Users demonstrate proactive engagement in managing their health, consistently performing health-related tasks, with a thorough understanding of their condition and a commitment to long-term well-being. & \\

\midrule

\addlinespace

\textbf{Cultural differences} & \makecell[l]{Individualism vs.\\collectivism} &
\textit{Individualistic} cultures may prioritise personalisation and user autonomy, whereas \textit{collectivist} cultures emphasize community-focused features, shared decision-making, or caregiver involvement. &
\textit{Collectivism}: His children visit him on weekends, and he regularly spends time at the nearby RSL club with fellow veterans. These visits provide a chance to share stories and discuss health-related topics, including their experiences with mHealth apps. \\

 & \makecell[l]{Long-term vs.\\short-term orientation} &
\textit{Long-term-oriented} users prefer apps that help build good habits over time and focus on future benefits, while \textit{short-term-oriented} users enjoy apps that offer quick results, instant rewards, and immediate health improvements. & \\

\midrule

\addlinespace

\textbf{Level of dependency} & Independent &
Independent users rely on mHealth apps as supplementary tools and require minimal external support, prioritising features that support autonomy, flexibility, and self-directed use. This facet spans functional, caregiver-mediated, and clinician-mediated dependence. &
\textit{Moderate dependence}: Daniel is mostly independent but relies on a home care assistant for weekly meal preparation and his children for help understanding health information and making informed decisions. \\

 & Moderate dependence &
Users rely on caregivers for specific health or daily tasks but manage others independently, using mHealth apps as a shared resource with caregiver support or simplified functionality to aid health management. & \\

 & Fully dependent &
Users rely on caregivers for most daily and health-related tasks, requiring them to share most or all health information to support informed decision-making. & \\

\midrule

\addlinespace

\textbf{Trust in mHealth apps} & Low trust &
Users have limited confidence in mHealth apps, questioning their ability to meet needs or provide reliable support, leading to hesitation or reluctance in adoption. &
\textit{Moderate trust}: Initially skeptical about mHealth apps, Daniel installed one after positive feedback from friends at the RSL club. \\

 & Moderate trust &
Users have a cautious belief in mHealth apps, recognising potential benefits but remaining unsure about their reliability or effectiveness in addressing all health-related needs. & \\

 & High trust &
Users have strong confidence in mHealth apps, expecting them to reliably support their health goals and effectively address their needs, which drives active engagement and adoption. & \\

\midrule

\addlinespace

\textbf{Users’ involvement in mHealth apps} & Passive involvement &
Users engage with mHealth apps in a limited or indirect manner, relying on automated processes or external inputs and demonstrating minimal willingness to interact with or actively control the system. &
\textit{Passive involvement}: As a newcomer to mHealth apps, Daniel has limited interest in exploring its features. His skepticism and lack of familiarity reduce his motivation to engage with the app actively. \\

 & Active involvement &
Users take an active role in engaging with mHealth apps, directly interacting with and managing the system, showing a strong willingness to explore its features and understand its functionality. & \\

\end{longtable}
}

\color{darkblue}
ChroniUXMag was synthesised by consolidating the qualitative categories that emerged from our STGT analysis (focus groups and interviews) and triangulating them with survey findings into a small set of higher-level facets. To further examine whether these facets reflect issues articulated in real-world mHealth use, we conducted an exploratory analysis of user reviews from widely used diabetes management applications, treating the review data as an additional grounding source rather than as a framework evaluation step. For each facet, we (i) defined its conceptual scope, (ii) specified representative \,\textit{ranges} (i.e., plausible values/states observed or implied in the data) that capture meaningful variation in chronic-disease mHealth use, and (iii) documented indicative evidence (quotes, survey patterns, and/or review themes) supporting the facet and its range endpoints.

\color{darkblue}
The remainder of this section shows how these facets are turned into reusable artefacts for requirements work: (1) personas that instantiate facet combinations, and (2) an adapted cognitive walkthrough form that prompts evaluators to explicitly ground their judgements in the facets.

\color{darkgreen}
\subsection{\rrev{ChroniUXMag Personas (Framework Artefact)}}
This section illustrates how ChroniUXMag facets are operationalised through the construction of example personas. The purpose is not to evaluate personas as an outcome in themselves, but to demonstrate how empirically derived facets can be translated into concrete, reusable artefacts that support requirements elicitation and analysis in mHealth contexts. The personas presented here serve as representative examples, showing how combinations of ChroniUXMag facets capture heterogeneous user situations and make inclusivity-relevant considerations explicit during downstream design and evaluation activities. From an RE perspective, personas are useful not merely as UX communication devices, but as lightweight artefacts for representing stakeholder differences, exposing hidden assumptions about users, and informing requirement refinement. Prior software engineering work has shown that personas and related user-centred artefacts can support requirements analysis by helping teams reason about user needs, interaction contexts, and design trade-offs \cite{karolita2023use,Lop2014,Tek2017,lopes2018applying}. In this way, ChroniUXMag personas function as a bridge between UX/HCI insights and RE practice: they translate qualitative and quantitative evidence about users into structured representations that can guide requirement discussions, prioritisation, and validation.

\color{black}
ChroniUXMag defines facet ranges representing health situations and stages in chronic disease management, which informed the creation of personas reflecting varied user needs and app interactions. Similar to GenderMag \cite{burnett2016gendermag}, which employs personas to uncover cognitive barriers in software use, our personas help identify variations in user needs and experiences within mHealth apps. Our example personas, available in the supplementary materials \footnote{\url{https://doi.org/10.5281/zenodo.17292487}}, provide practical insights. To ensure that personas effectively inform the requirement elicitation process, the selected \rrev{facets} must accurately represent real-world user behaviours and acknowledge the \textbf{interdependencies} between different factors \cite{nielsen2017who}. For example, \rrev{facets} such as \hconcep{\textit{health situation}}, \hconcep{\textit{level of dependency}}, and \hconcep{\textit{user involvement}} are interdependent, shaping the extent to which users rely on caregivers for mHealth app engagement. Patients with moderate to severe chronic diseases often show passive \hconcep{\textit{involvement}}, rely on caregivers for app interactions, and share most health-related information, indicating full dependency.

\begin{figure}
    \centering
    \includegraphics[width=0.9\linewidth]{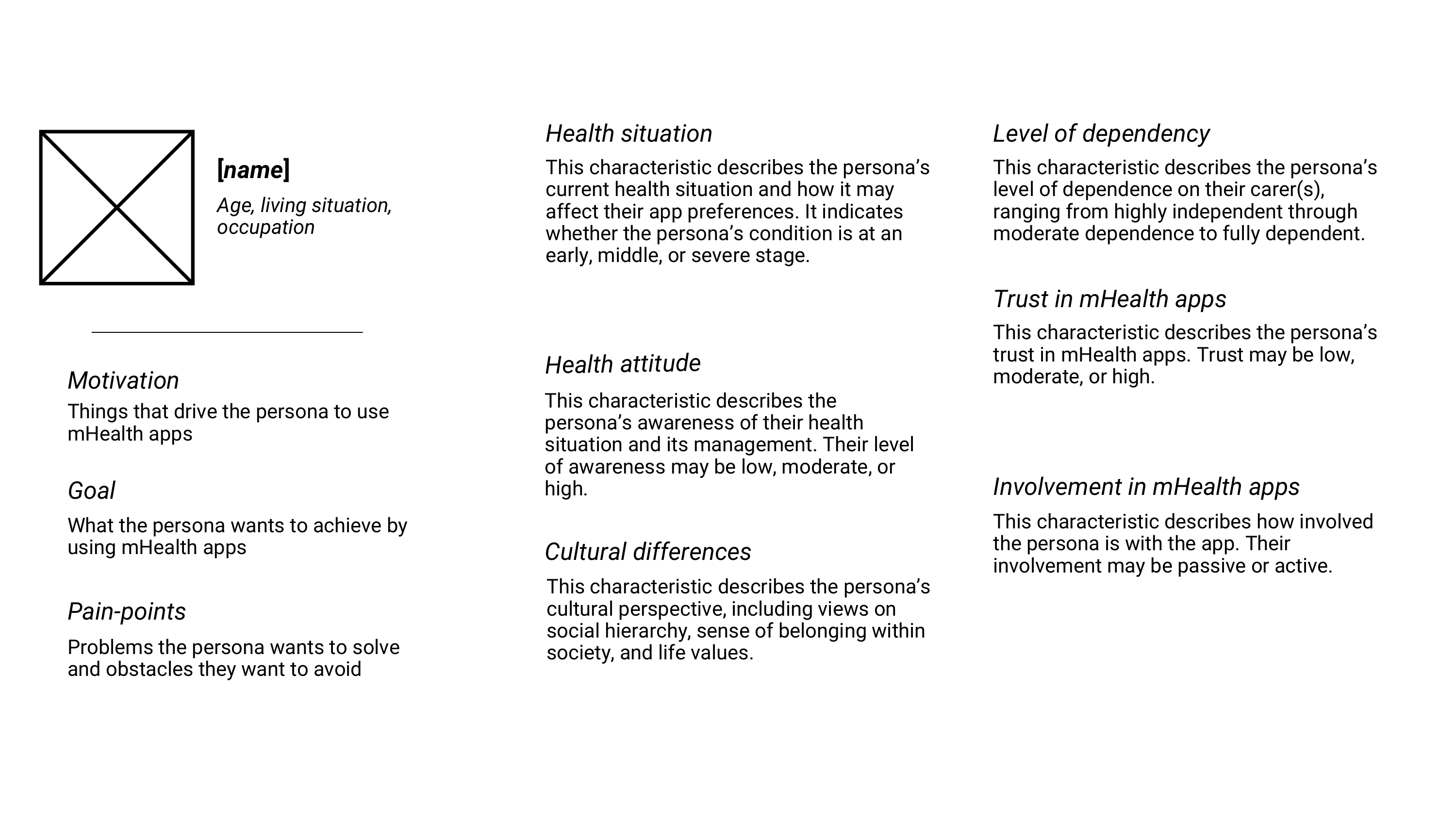}
    \caption{ChroniUXMag Persona Template}
    \label{fig:persona}
\end{figure}

\color{black}
To operationalise the ChroniUXMag framework, we developed a structured persona template (Figure \ref{fig:persona}) that integrates traditional persona elements with facets specific to chronic disease management and mHealth use. The template begins with demographic and contextual details, such as age, living situation, and occupation, along with motivations, goals, and pain points that capture the reasons to engage with mHealth apps and the obstacles users seek to overcome. It then incorporates six facets derived from ChroniUXMag. By populating these elements, practitioners and researchers can construct personas that reflect diverse user needs and experiences. These personas can then be applied in design and evaluation activities. In this way, the persona template serves as a practical tool for ensuring inclusivity and accessibility are embedded into the design of mHealth apps.

\color{black}

\color{black}

\color{black}

\color{black}

\color{black}
\subsection{\rrev{Adapted Cognitive Walkthrough Form}}
\begin{figure}[h]
        \centering
        \includegraphics[width=0.75\linewidth]{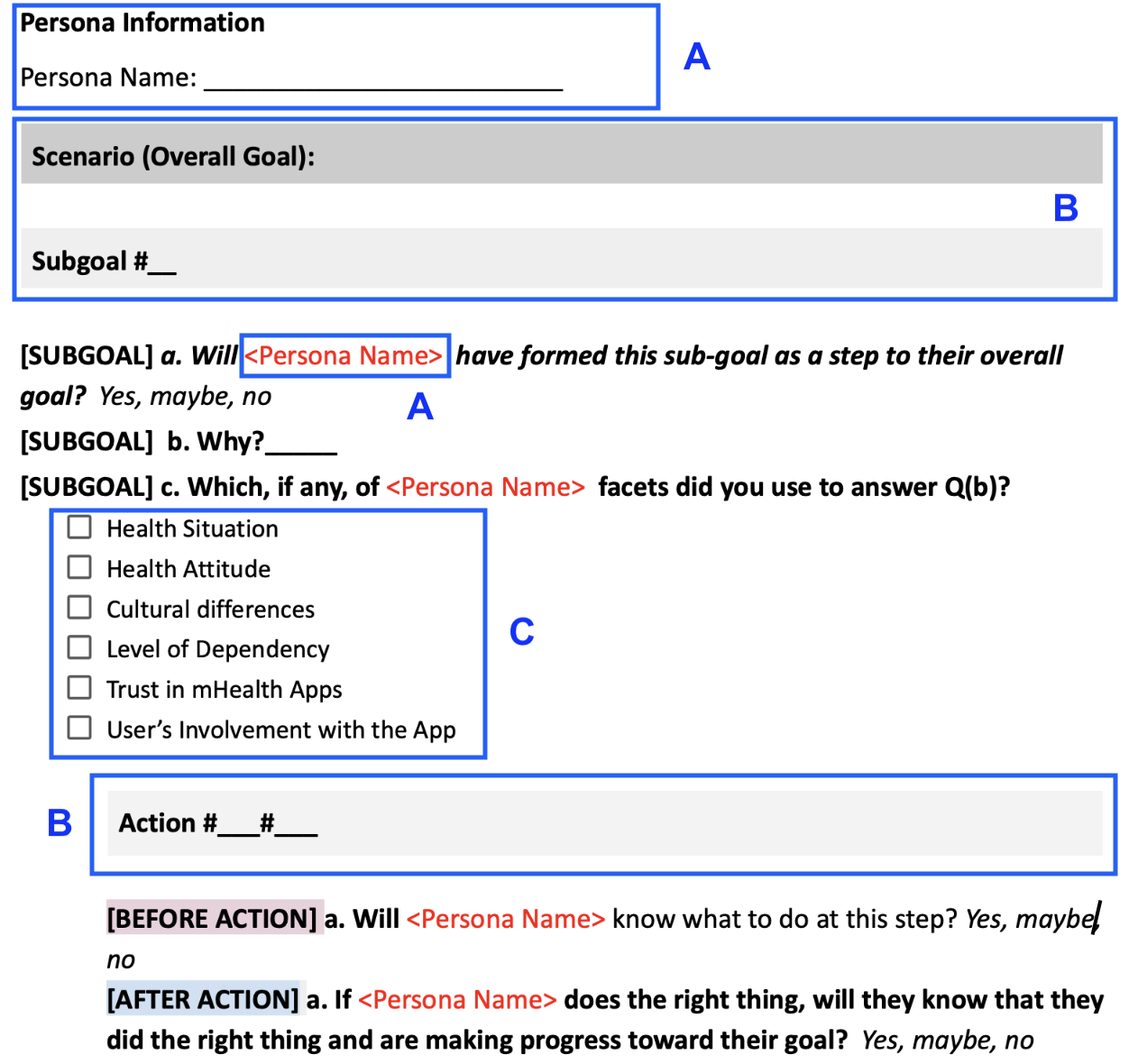}
        \caption{ChroniUXMag specialisation of a Cognitive Walkthrough form}
        \label{fig:CW}
    \end{figure}

\color{darkblue}
This section describes how ChroniUXMag facets are embedded into a specialised cognitive walkthrough to support inclusivity-focused evaluation of mHealth applications. The walkthrough adaptation is presented as a method for applying the framework in practice, rather than as a standalone evaluation study. Through this process, we illustrate how persona facets systematically guide evaluators’ reasoning and help surface both general usability issues and inclusivity-related concerns. ChroniUXMag specializes the cognitive walkthrough by integrating personas that embody six facets. These facets shape how users with chronic conditions perceive, interact with, and place trust in mHealth applications. During the walkthrough, practitioners adopt the perspective of a selected persona and work through the evaluation template by answering structured questions (see Figure \ref{fig:CW}(A)). Each overall goal is decomposed into subgoals, which are further broken down into concrete actions (see Figure \ref{fig:CW}(B)). For every action step, evaluators consider both the before-action and after-action perspectives. Specifically, they ask: (a) Before action: Will the persona know what to do?; (a) After action: If the persona performs the action, will they recognise that it was correct and that progress has been made toward their goal?; (b) Why?; and (c) Which, if any, of the persona’s facets informed the answer to (b)? To support responses to question (c), the template provides a structured list of the persona’s facets, enabling evaluators to explicitly link their reasoning to facets (see Figure \ref{fig:CW}(C)).

\color{darkblue}
Issues identified during the walkthrough are then classified as either general usability issues or inclusivity issues tied to specific ChroniUXMag facets. For example, a button labelled in overly technical language might create a general usability problem for most users, whereas a small text size or unexplained data-sharing prompt may disproportionately affect users with low digital literacy, heightened privacy concerns, or dependency on caregiver support. By embedding ChroniUXMag facets into the cognitive walkthrough process, our approach extends the method beyond traditional usability to support early requirements analysis in chronic-disease mHealth applications. In this paper, the walkthrough is not positioned only as a UX evaluation instrument; rather, it is used as a structured mechanism for revealing requirement gaps, hidden assumptions, and design constraints that may otherwise remain implicit. Findings from the walkthrough can therefore be interpreted in RE terms, for example as candidate inclusivity requirements, refinements to existing requirements, or evidence that a proposed design does not adequately satisfy the needs of particular stakeholder groups. This is the main point of intersection between the UX/HCI origins of the method and its contribution to RE.

\color{darkgreen}
\section{\rrev{Illustrative Use of ChroniUXMag}} \label{sec:apply-chroniuxmag}
\begin{figure}
    \centering
    \includegraphics[width=1\linewidth]{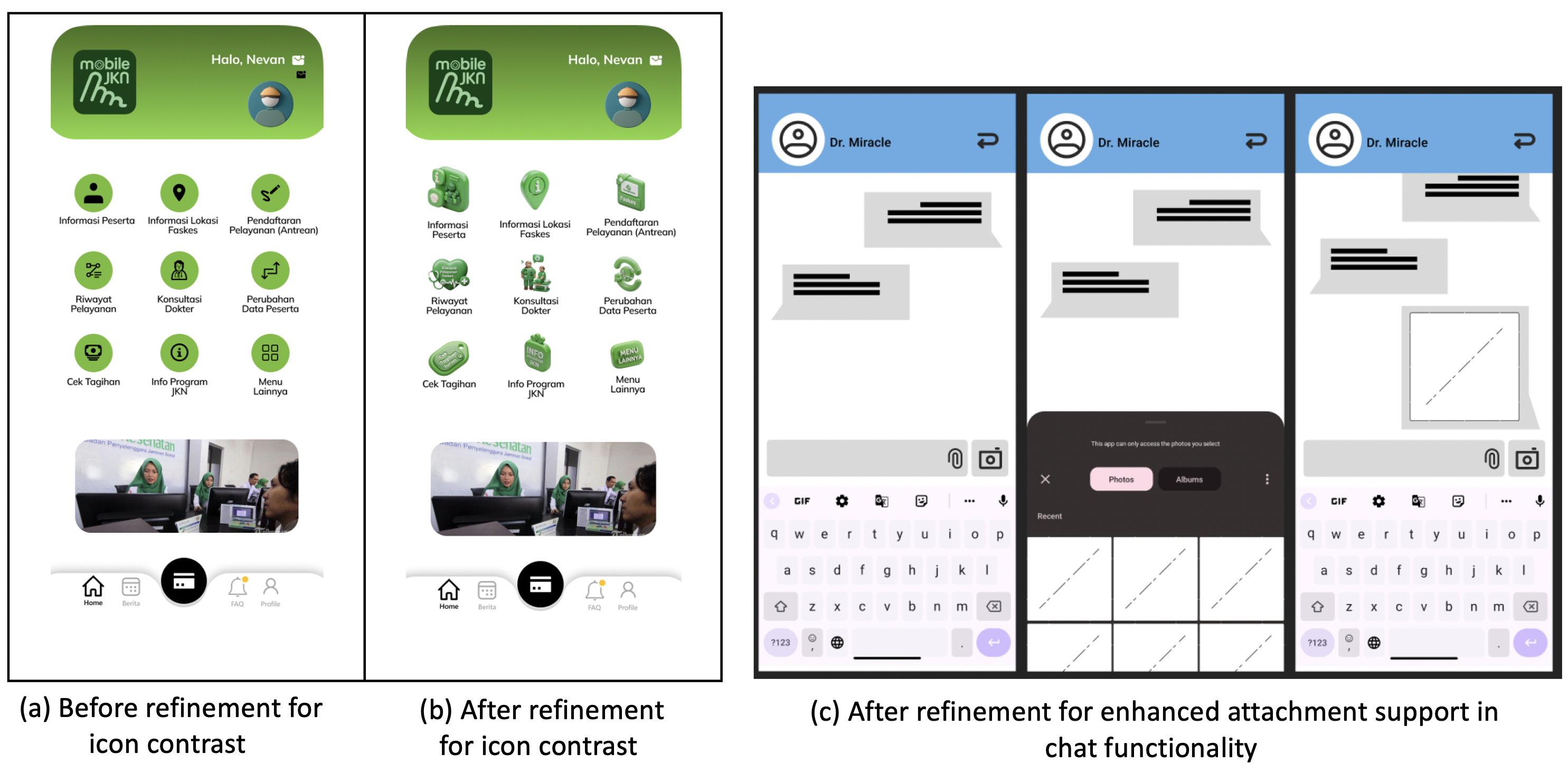}
    \caption{Example prototype refinements informed by ChroniUXMag-guided cognitive walkthrough, including improved icon contrast and enhanced attachment support in chat functionality}
    \label{fig:cwexample}
\end{figure}

To illustrate the use of ChroniUXMag, we briefly report representative issues from an ongoing course-based software engineering class where student teams used ChroniUXMag personas and the walkthrough form to develop and evaluate an mHealth solution. These walkthroughs were conducted as part of a course assignment and are reported here for illustration rather than as a formal evaluation. Using a ChroniUXMag persona and CW template, evaluators identified several recurring issues that were not purely general usability problems but were explicitly linked to inclusivity facets. Using a ChroniUXMag persona representing a user with reduced stamina and lower tolerance for cognitive load, evaluators identified that insufficient visual contrast between feature icons and the background increased navigation effort, particularly in light mode. This issue was not prioritized when using a general, digital-native persona. The finding led to a prototype refinement that increased icon contrast to improve visual clarity (See Figure \ref{fig:cwexample}). In a second example, the walkthrough revealed that a non-functional attachment feature in a chat-based consultation tool prevented users from sharing images or documents, a critical barrier for users with higher dependency on remote care and caregiver-mediated interactions. This issue was explicitly linked to ChroniUXMag facets related to dependency and trust, and subsequently informed the redesign of the chat interface to support image and document uploads (See Figure \ref{fig:cwexample}). These examples demonstrate how ChroniUXMag helps evaluators surface inclusivity-specific concerns and translate them into actionable design refinements, without requiring a full empirical evaluation of the framework.

\color{black}

\color{black}
\section{Threat to validity} \label{threat to validity}

\color{darkblue}
\textbf{Construct Validity:} Our methodology combines literature review, qualitative and quantitative studies, and persona-driven evaluation. While this triangulation enhances robustness, several risks remain. Personas, while useful for capturing diverse user profiles, are abstractions and cannot fully represent the complexity of chronic disease experiences. The development of facets relied on researcher interpretation, which introduces subjectivity and may oversimplify nuanced health and accessibility needs. To mitigate this, personas were grounded in empirical data derived from qualitative analysis using Socio-Technical Grounded Theory (STGT), with explicit memoing and constant comparison used to trace how categories and facets emerged from participant accounts. ChroniUXMag builds on existing persona-based inclusivity approaches, particularly GenderMag and InclusiveMag. A potential threat is that alternative RE-oriented frameworks could have been selected. We chose these approaches because they provide a well-established, empirically grounded mechanism for operationalising inclusivity concerns through personas and walkthrough-based evaluation, which aligns with our goal of supporting requirements elicitation and early analysis rather than late-stage accessibility compliance. Another threat arises from the heterogeneity of chronic diseases represented in our participant pool, as the study did not focus on a single condition but included participants with a wide range of chronic diseases, including cardiometabolic, respiratory, immune-related, and mental health conditions, often in combination. This heterogeneity may obscure condition-specific requirements. However, this was a deliberate design choice aligned with our research goal of identifying cross-cutting inclusivity facets that recur across chronic disease contexts rather than disease-specific clinical requirements. To mitigate this threat, we focused our analysis on user-reported experiences, challenges, and perceptions related to mHealth usage, rather than clinical characteristics of individual diseases.In addition, the framework assessment based on user reviews was preliminary. This analysis was not a full evaluation of ChroniUXMag’s effectiveness as a requirements engineering method, but an exploratory relevance check to see whether the identified facets align with concerns expressed in real-world mHealth app use. We explicitly limit our claims accordingly and position comprehensive evaluation of the framework’s structuredness and practical utility as future work.

\color{darkblue}
\textbf{Internal Validity:} Internal validity concerns whether the observed findings can be attributed to the study procedures rather than unintended bias or methodological artefacts. The process of mapping qualitative insights into structured facets and then abstracting them into persona attributes introduces potential bias. Although Socio-Technical Grounded Theory (STGT) was applied systematically, researcher judgement played a role in coding and memoing, which could affect objectivity. Survey data provided complementary insights, but self-reported measures may be prone to recall or social desirability bias. Finally, our customized cognitive walkthrough builds on GenderMag’s method; however, its evaluation in this study was limited, meaning that any observed inclusivity issues may be partially shaped by the walkthrough structure itself.

\color{darkblue}
\textbf{External Validity:}  The empirical studies were conducted with a specific subset of participants (Table \ref{tab:demographics}), largely drawn from support groups, networks, and online forums. While these participants offered diverse perspectives, they may not represent the full demographic, cultural, and health-condition spectrum of the broader chronic disease population. Although participants were recruited from multiple countries, the sample exhibits asymmetries in nationality and gender distribution, with a higher proportion of participants from Australia and a gender imbalance. These factors may limit the generalisability of findings to other populations or cultural contexts. We mitigate this threat by reporting demographic distributions transparently and by avoiding claims of universal applicability. Moreover, our case study focused on four diabetes management applications, limiting the applicability of the findings to other chronic conditions. Restricting user review analysis to English-language reviews also risks omitting cultural and linguistic nuances in how inclusivity challenges are expressed.

\color{darkblue}
\textbf{Conclusion Validity:}
Conclusion validity concerns the strength and reliability of inferences drawn from the data. The survey analysis was constrained by sample size, particularly in subgroup comparisons, which may reduce statistical power and limit confidence in observed trends. Similarly, the exploratory user review analysis required interpretive mapping of unstructured comments to ChroniUXMag facets, as reviews were not originally written with these facets in mind. This introduces potential ambiguity in linking user statements to specific inclusivity \rrev{facets} and limits the extent to which causal relationships can be inferred between user characteristics and identified mHealth challenges. Accordingly, we interpret quantitative patterns conservatively and position review-based findings as indicative rather than confirmatory.

\color{black}
\section{Future Work} \label{future work}

\color{darkblue}
This study introduced ChroniUXMag as a structured framework for capturing inclusivity concerns in mHealth applications, grounded in empirical data from individuals living with chronic conditions and informed by established persona-based inclusivity methods such as InclusiveMag and GenderMag. The present work focused on the \textit{Scope} and \textit{Derive} stages of InclusiveMag, resulting in the identification and structuring of inclusivity-related facets. The \textit{Apply} stage, which concerns systematic use and evaluation in design practice, remains a key direction for future research.

\color{darkblue}
Future work will operationalise the \textit{Apply} stage by examining how ChroniUXMag is used by practitioners in real-world and educational design settings. Planned studies include workshops and design activities involving software developers, UX designers, clinicians, and accessibility specialists. Participants will apply ChroniUXMag personas and the adapted cognitive walkthrough form to concrete mHealth scenarios, generating inclusivity-related findings at the feature, workflow, and socio-technical levels.

\color{darkblue}
These studies will enable evaluation of ChroniUXMag against explicit criteria that address current validity limitations. In particular, we will examine (i) \textit{scalability}, by assessing how the framework performs across different project sizes and chronic disease contexts; (ii) \textit{inter-rater reliability}, by analysing the consistency of inclusivity issues identified by different evaluators using the same personas and walkthrough materials; and (iii) \textit{practical usefulness}, by investigating how the identified issues inform design decisions and refinements. Longitudinal studies will further explore how ChroniUXMag-informed requirements influence subsequent design iterations over time.

\color{darkblue}
By embedding ChroniUXMag into practitioner-led evaluation and design processes, future work aims to provide a more rigorous assessment of its applicability, limitations, and contribution to inclusive requirements engineering for mHealth systems.

\color{black}

\color{darkblue}
\section{Conclusion} \label{conclusion}
This paper introduced ChroniUXMag, a framework for eliciting and analysing inclusivity requirements in mHealth applications for chronic disease management. Informed by principles from InclusiveMag and GenderMag, ChroniUXMag currently encompasses two stages, \textit{Scope} and \textit{Derive}, through which literature-based insights and empirical evidence are synthesised to identify facets that reflect the heterogeneous and evolving needs of people living with chronic diseases. The findings demonstrate how these empirically grounded facets enable systematic identification of both usability and inclusivity concerns that are often overlooked in conventional evaluation approaches.

\color{darkblue}
From a requirements engineering perspective, this work contributes a structured, facet-driven and persona-based mechanism for embedding inclusivity into requirements elicitation and early analysis. Although ChroniUXMag is informed by UX/HCI methods such as InclusiveMag, GenderMag, personas, and cognitive walkthroughs, its contribution in this paper is not limited to interface evaluation. Rather, we show how these methods can be repurposed as RE artefacts and analysis mechanisms: empirical user insights are translated into personas and facets, and walkthrough findings are interpreted as requirement-relevant evidence about usability barriers, accessibility needs, trust concerns, and caregiver-mediated interactions. By making these concerns explicit, analysable, and traceable, ChroniUXMag helps connect UX/HCI-informed user understanding with core RE activities in chronic disease mHealth. In doing so, this research advances the integration of inclusivity into RE practice, not as a downstream usability check, but as a systematic component of requirements elicitation, analysis, and refinement.

\color{black}

\color{black}

\color{black}

\color{black}

\color{black}

\color{black}

\color{black}

\color{black}
\section*{Acknowledgement}
Wang and Grundy are supported by ARC Laureate Fellowship FL190100035.

\bibliographystyle{IEEEtran}

\end{document}